\documentclass{sig-alternate}
\usepackage[ruled,vlined,linesnumbered]{algorithm2e}
\usepackage{amsmath,amssymb}
\usepackage{epsfig}
\usepackage{color}
\usepackage{url}
\usepackage{varioref}
\newcommand{\subparagraph}{} %introduced for titlesec package
\usepackage[compact]{titlesec} %for squeezing. refer titlespacing in squeeze
\usepackage{setspace} %for reducing the space in bibliography
\usepackage{pifont} %for the markers on the author's affiliation
\usepackage[font=bf]{caption}

\newcommand{\theoremname}{Theorem}
\newcounter{thm}
\newenvironment{thm}{%
  \par\medskip\refstepcounter{thm}%
  \noindent\textbf{Theorem \thethm:}\quad}{\par\medskip}
\labelformat{thm}{\theoremname~#1}

\newcommand{\definitionname}{Definition}
\newcounter{defn}

\labelformat{defn}{\definitionname~#1}

\newcommand{\lemmaname}{Lemma}
\newcounter{lemma}

\labelformat{lemma}{\lemmaname~#1}

\newcommand{\corollaryname}{Corollary}
\newcounter{corollary}

\labelformat{corollary}{\corollaryname~#1}

\DeclareMathOperator*{\argmin}{arg\,min}

\def\norm#1{{\left\|\,#1\,\right\|}}

\def\norm #1{\|#1\|}

\def\argmin{\mathop{\rm arg\,min}}

\newcommand{\algofull}[0]{{Text Outliers using Nonnegative Matrix Factorization}}
\newcommand{\algo}[0]{{TONMF}}

\newcommand{\ramki}[1]{#1}

\begin{document}

\title{Outlier Detection for Text Data : An Extended Version}

\numberofauthors{4}
%
%\author{
%\alignauthor Ramakrishnan Kannan \textsuperscript{\ddag}\\
%%\affaddr{Georgia Institute of Technology}\\
%%\affaddr{Atlanta, GA, USA}\\
%\email{rkannan@gatech.edu}
%%
%\alignauthor Hyenkyun Woo \textsuperscript{\ding{169}}\\
%%\affaddr{Korea Institute for Advanced Study}\\
%%\affaddr{Seoul, Korea}\\
%\email{hyenkyun@kias.re.kr}
%%
%\alignauthor Charu C. Aggarwal \textsuperscript{\ding{86}}\\
%%\affaddr{IBM T. J. Watson Research Center}\\
%%\affaddr{Hawthorne, NY, USA}\\
%\email{charu@us.ibm.com}
%%
%%\and
%\alignauthor Haesun Park\textsuperscript{\ddag}\\
%%\affaddr{Georgia Institute of Technology}\\
%%\affaddr{Atlanta, GA, USA}\\
%\email{hpark@cc.gatech.edu}
%\end{tabular}\newline\begin{tabular}{c}
%\affaddr{\textsuperscript{\ddag}Georgia Institute of Technology,Atlanta, GA, USA}\\
%\affaddr{\textsuperscript{\ding{169}}Korea Institute for Advanced Study, Seoul, Korea}\\
%\affaddr{\textsuperscript{\ding{86}}IBM T. J. Watson Research Center, Hawthorne, NY, USA}\\
%}

\author{
\alignauthor Ramakrishnan Kannan\\
\affaddr{Oak Ridge National Laboratory}\\
\affaddr{Oak Ridge, TN, USA}\\
\email{kannanr@ornl.gov}
\alignauthor Hyenkyun Woo \\
\affaddr{Korea University of Technology and Education}\\
\affaddr{Republic of Korea}\\
\email{hyenkyun@koreatech.ac.kr}
\alignauthor Charu C. Aggarwal\\
\affaddr{IBM T. J. Watson Research Center}\\
\affaddr{Yorktown Heights, NY, USA}\\
\email{charu@us.ibm.com}
\and
\alignauthor Haesun Park\\
\affaddr{Georgia Institute of Technology}\\
\affaddr{Atlanta, GA, USA}\\
\email{hpark@cc.gatech.edu}
}

\maketitle

\begin{abstract}

The problem of outlier detection is extremely challenging in many
domains such as text, in which the attribute values are typically
non-negative, and most values are zero. In such cases, it often
becomes difficult to separate the outliers from the  natural
variations in the patterns in the underlying data.  In this paper,
we present a matrix factorization method, which is naturally able to
distinguish the anomalies with the use of low rank approximations of
the underlying data.  Our iterative algorithm TONMF is based on
block coordinate descent (BCD) framework. We define blocks over the
term-document matrix such that the function becomes solvable. Given
most recently updated values of other matrix blocks, we always
update one block at a time to its optimal. Our approach has
significant advantages over traditional methods for text outlier
detection. Finally, we present experimental results illustrating the
effectiveness of our method over competing methods.

\end{abstract}

\section{Introduction}
\label{sec:motivation}
 The problem of outlier detection is that of finding data points
 which are unusually different from the  rest of the data set.
 Such outliers are also variously referred to as anomalies, deviants,
 discordants or abnormalities in the data. Since outliers  correspond to
 unusual observations, they are often of interest to the analyst in
 finding interesting anomalies in the underlying generating process.
 \pagebreak
 The problem of
 outlier analysis is applicable to a wide variety of domains such as
 machine monitoring, financial markets, environmental modeling and
 social network analysis. 
Correspondingly, the problem has been studied in the context of
different data types which arise in these domains, such as
multidimensional data, spatial data, and discrete sequences.
Numerous books and surveys have been written on the problem
 of outlier detection \cite{outlierbook,chandola,chandola2,hawkins}.

In this paper, we will study the problem of text outlier analysis.
The problem of text outlier analysis has become increasingly
important because of the greater prevalence of web-centric and
social media applications, which are rich in text data. Some
important applications of text outlier analysis are as follows:
\begin{itemize}
\item {\em Web Site Management:} An unusual page from a set of
articles in a web site may be flagged as an outlier. The knowledge
of such outliers may be used for web site management.
\item  {\em Sparse High Dimensional Data:} While the methods
discussed in this paper have text applications in mind, they can be
used for other sparse high dimensional domains. For example, such
methods can be used for market basket data sets. Unusual
transactions  may sometimes provide an idea of fraudulent behaviour.
\item {\em News Article Management:}   It is often desirable to
determine unusual news article from a collection of news documents. An
unusual news from a group of articles may be flagged as an
interesting outlier.
\end{itemize}
While text is an extremely important domain from the perspective of
outlier analysis, there are surprisingly few methods which are {\em
specifically focused}  on this domain, even though many generic
methods such as distance-based methods can be easily adapted to this
domain \cite{knorr,rama}, and are often used for text outlier
analysis. Domains such as text are particularly challenging for the
 problem of outlier analysis, because of their sparse high
 dimensional nature, in which only a small fraction of the words
 take on non-zero values.  Furthermore, many  words in a document
 may  be  topically irrelevant to the context of the document and add to the noise in the
 distance computations. For example, the word ``{\em Jaguar}'' may correspond
 to a car, or a cat depending on the context of the document.
     In
 particular, the significance of a word can be interpreted only in
 terms of the structure of the data within the context of a
 particular data locality.  As a result, document-to-document similarity
 measures often lose their robustness. Thus, commonly used
 outlier analysis methods for multidimensional data, such as distance-based methods, are not
 particularly effective for text data. Our experiments also validate this observation. 

In this paper, we will
  use  non-negative matrix factorization (NMF) methods to  address the
  aforementioned challenges in text anomaly detection. One
  advantage of matrix factorization methods is that they decompose  the term-document
 structure of the underlying corpus into a set of semantic term clusters and document clusters.
 The semantic nature of this decomposition provides the  context in
 which a document may be interpreted for outlier analysis.
 Thus, documents can be decomposed into word clusters, and words are
 decomposed into document clusters with a low-rank\footnote{In this paper, we use  the terms  ``low rank
approximation'' and ``matrix factorization'' interchangeably.
Similarly, we used the terms  ``anomalies'' and ``outliers''
interchangeably.} approximation.
 Outliers are therefore defined as data points which cannot be
 naturally expressed in terms of this decomposition.
   By using carefully chosen model formulations,
 one can further sharpen the matrix-factorization method to reveal
 document-centric outliers. One  challenge in this case, is that the
 design of a matrix factorization approach, which is optimized to
 anomaly detection, results in a non-standard formulation.
 Therefore, we will design an optimization solution for this model.
 The NMF model also has the advantage of providing better
 interpretability, and it can also provide insights into why a
 document should be considered an outlier.   We present extensive
experimental results  on many   data sets, and compare against a
variety of baseline methods. We show significant improvements
achieved by the approach over a variety of other methods.

This paper is organized as follows. The remainder of this section
discusses the related work.  Section \ref{sec:model} introduces the
model for outlier analysis. The algorithm to solve  this model is
provided in section \ref{sec:algorithm}. Section
\ref{sec:experimentation} provides the experimental results. The
conclusions and summary are contained in section
\ref{sec:conclusion}. Our code can be downloaded from 
\url{https://github.com/ramkikannan/outliernmf} and 
tried with any text dataset.

\subsection{Related Work} \label{sec:survey}

The outlier analysis problem has been studied extensively in the
literature \cite{outlierbook,chandola,hawkins}.  Numerous algorithms
have been proposed in the literature for outlier detection of
conventional multidimensional data \cite{hd,lof,knorr,rama}. The key
methods, which are used  frequently for outlier analysis include
distance-based methods \cite{knorr,rama}, density-based methods
\cite{lof}, and subspace methods \cite{hd,keller,laz,muller,zimek}.
In distance-based methods, data points are declared outliers, when
they are situated far away from  the dense regions in the underlying
data. Typically, indexing or other summarization schemes may be used
in order to improve the efficiency of the approach. In density-based
methods \cite{lof},   data points with low local density with
respect to the remaining points are declared outliers. In addition,
a number of subspace methods \cite{hd,keller,laz,muller,zimek} have
been proposed recently, in which outliers are defined on the basis
of subspace behavior of the underlying data.

Most of the traditional multidimensional
methods \cite{chandola,outlierbook} can also be extended to text
data, though they are not particularly suited to the latter. Some
methods have been designed for outlier detection with matrix
factorization in network data sets \cite{tong}, that
are not applicable to text data. Text data is uniquely
difficult because of its sparse and high dimensional nature.  As a
result, many of the outliers detected using conventional methods may
simply correspond to noisy text segments. Therefore, careful
modeling is required with the use of matrix factorization methods.

Over the last decade, Non-negative Matrix Factorization (NMF) has
emerged as another important low rank approximation technique, where
the low-rank factor matrices are constrained to have only
non-negative elements. Lee and Seung \cite{Lee1999} introduced a
multiplicative update based low rank approximation with non-negative
factors to overcome the challenges of truncated SVD. Subsequent to
this work, NMF has received enormous attention and has been
successfully applied to a broad range of important problems in areas
including computer vision, community detection in
social networks, visualization, recommender systems bioinformatics,
etc. In spite of broad range of applications,
NMF's literature in text domain is scarce. Xu {\em et. al.}
\cite{Xu2003} experimented with NMF for document clustering instead
of SVD based Latent Semantic Indexing (LSI). Other than applications
of NMF in the  text domain, Gaussier and Goutte \cite{Gaussier2005}
established the equivalence between NMF and pLSA. Similarly, Ding
{\em et. al.} \cite{Ding2006}  explained the equivalence between NMF
and pLSI.

In this paper, we use an NMF approach for concise modelling of the
patterns, the background, and the anomalies in the underlying data.
It should be pointed out that NMF is similar to the generative
models of text such as pLSI and LDA \cite{Gaussier2005}
\cite{Ding2006} \cite{Singh2008}, though NMF often provides better
interpretability. Our important challenge is to model the outliers
along with the low rank space of the input matrix. We identified
$\ell_{1,2}$-norm as an appropriate approach for factorization in
outlier analysis. Recently, the researchers have used  $\ell_{2,1}$-norm in their models to solve various problems, though the
corresponding solution techniques are not easily generalizable to
the $\ell_{1,2}$-norm. Yang et.al., \cite{Yang2011}, under the
assumption that the class label of input data can be predicted by a
linear classifier, incorporate discriminative analysis and
$\ell_{2,1}$-norm minimization into a joint framework for
unsupervised feature selection problem. Similarly, Liu {\em et al}
\cite{Liu2009}, solve $\ell_{2,1}$-norm regularized regression model
for joint feature selection from multiple tasks. They also propose
to use Nesterov's method to solve the optimization problem with
non-smooth $\ell_{2,1}$-norm regularization. Also, Kong {\em et al}
\cite{Kong2011} propose a robust formulation of NMF using
$\ell_{2,1}$-norm loss function for data with noises.
\subsection{Our Contributions}
Text data is uniquely challenging to outlier detection both because
of its sparsity and high dimensional nature.  Given the relevant
literature for NMF and text outliers, we propose the first approach
to detect outliers in text data using non-negative matrix
factorization. We extend the fact that NMF is similar to pLSI and
LDA generative models and model the outliers using the
$\ell_{1,2}$-norm.  This particular formulation of NMF is
non-standard, and requires careful design of optimization methods to
solve the problem. We solve the resulting optimization problem using
block coordinate descent technique. We also present extensive
experimental results both on text and other kinds of market basket
data sets. We show significant improvements achieved by the approach
over other baseline methods.

%\section{notations}
%\label{sec:notations}
\begin{table}
\begin{tabular}{ l | l }
\hline
Notation & Explanation\\ \hline
  $\mathbf{A}=[\mathbf{a}_1 \cdots \mathbf{a}_n] \in \mathbb{R}_+^{m \times n}$ & Document-word matrix  \\
  $m$ & Vocabulary size \\
  $n$ & Number of documents \\
  $\mathbf{Z} \in  \mathbb{R}^{m \times n}$ & Outlier matrix\\
  $r < rank(\mathbf{A})$ & Rank \\
  $\mathbf{W} \in \mathbb{R}_+^{m \times r}$ & Term-Topic matrix\\
  $\mathbf{H} \in \mathbb{R}_+^{r \times n}$&  Topic-Document matrix \\
  $\mathbf{A}^{(i)}$ & Matrix $\mathbf{A}$ from the $i^{th}$ iteration \\ 
  $\norm{\mathbf{A}}_{1,2}$ & $\sum_{i=1}^n\norm{\mathbf{a}_i}_{\ell_2}$ $\ell_{12}$-Norm where,\\  &  $\mathbf{a}_i \in \mathbb{R}^m$ is the $i$-th column of $\mathbf{A}$ \\
  \hline
  
\end{tabular}
\caption{Notations used in the paper} \label{table:notations}
\end{table}

\section{Matrix Factorization Model}
\label{sec:model} This section will  present the matrix
factorization model which is used for outlier detection. Before
discussing the model in detail, we present the notations and
definitions.  We represent the corpus of text documents as a bag of
words matrix.  A lowercase or uppercase letter such as $x$ or $X$,
is used to denote a scalar. A boldface lowercase letter, such as
$\mathbf{x}$, is used to denote a vector, and  a boldface uppercase
letter, such as $\mathbf{X}$, is used to denote a matrix. This is
consistent with what is commonly used in much of the data mining
literature. Indices typically start from $1$, unless otherwise
mentioned.  For a $\mathbf{X}$, $\mathbf{x}_{i}$ denotes its
$i^{th}$ column, $\mathbf{y}_j^\intercal$ denotes its $j^{th}$ row
and $x_{ij}$ or $X(i,j)$ or $(X)_{ij}$ denote its $(i,j)^{th}$
element.

For greater expressibility, we have also borrowed certain notations
from matrix manipulation scripts such as Matlab and Octave.  For
example, the notation $max(\mathbf{x})$ returns the maximal element
$x \in \mathbf{x}$ and $max(\mathbf{X})$ returns a vector of maximal
elements from each column $\mathbf{x} \in \mathbf{X}$. Similarly,
$\mathbf{X}(i,:)$ denotes the $i$-th row of the matrix and
$\mathbf{X}(:,i)$ for $i$-th column.  For the  reader's convenience,
the notations  used in the paper are summarized in  Table
\ref{table:notations}.

Let $\mathbf{A}$ be the matrix representing the underlying data. In
the context of a text collection, this corresponds to a
term-document matrix, where terms correspond to rows and documents
correspond to columns. In other words, $a_{ij}$ denotes the number
of times the term $i$ appears in document $j$.  Generally, we
can write $\mathbf{A}$ as follows:
\begin{equation}
\mathbf{A} = \mathbf{L_0} + \mathbf{Z_0}.
\end{equation}

\begin{figure}
\scalebox{0.34}[0.29]{\includegraphics{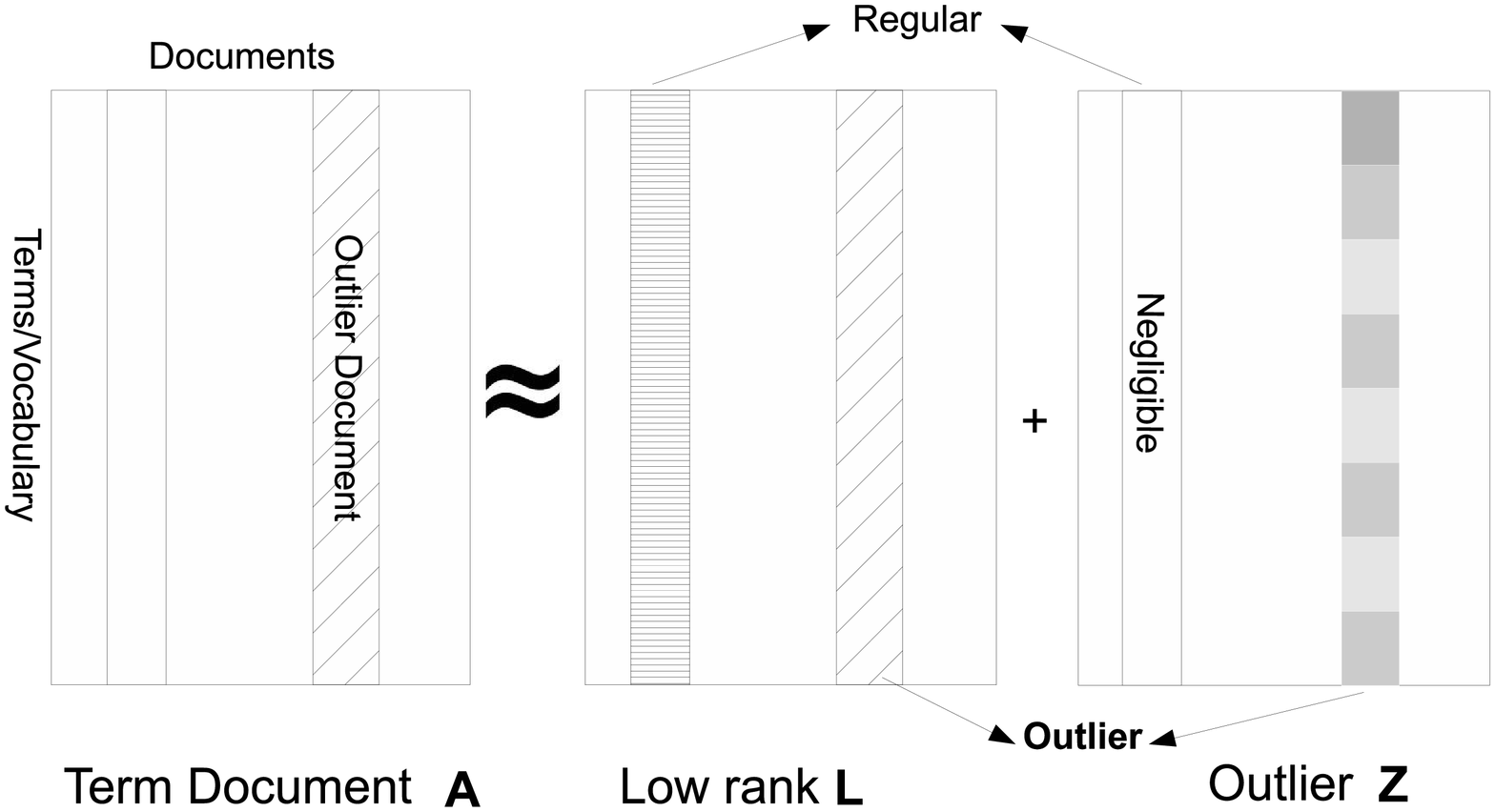}}
\caption{Text Outliers Using NMF} \label{fig:outlieroverview}
\end{figure}

Here, $\mathbf{L_0}$ is  a low rank matrix and $\mathbf{Z_0}$
represents the matrix of  outlier entries. Typically, the matrix
$\mathbf{L_0}$ represents the documents created by a lower rank
generative process (such as that modeled by pLSI), and the parts of
the documents that do not correspond to the generative process are
represented as part of the matrix $\mathbf{Z_0}$.   In real world
scenarios, the outlier matrix $\mathbf{Z_0}$  contains  entries
which are very close to zero, and  only a small number of entries
have {\em significantly} non-zero values. These significantly
nonzero entries are often present in only a small fraction of the
columns. Columns which are fully representable in terms of factors
are consistent with the low rank behavior of the data, and therefore
{\em not} outliers. The rank of $\mathbf{L}_0$ is not known in
advance, and it can be expressed in terms of its underlying factors.
$$
\mathbf{L}_0 \approx \mathbf{W}_0\mathbf{H}_0
$$
Here, the two matrices   have dimensions $\mathbf{W}_0 \in
\mathbb{R}^{m \times r}_+$, $\mathbf{H}_0 \in \mathbb{R}^{r \times
n}_+$, and $r \le rank(\mathbf{L}_0)$. The matrices $\mathbf{W}_0$
and  $\mathbf{H}_0$ are non-negative, and this provides
interpretability in terms of being able to express a document as a
non-negative linear combination of the relevant basis vectors, each
of which in itself can be considered a frequency-annotated bag of
words (topics) because of its non-negativity.  Specifically,
$\mathbf{H_0}$ corresponds to the coefficients for the basis matrix
$\mathbf{W_0}$. Intuitively, this corresponds to the case that every
document
 $\mathbf{a}_i$, is represented as the linear combination of
the $r$ topics. In cases, where this is {\em not} true, the document
is an outlier, and those  unrepresentable sections of the matrix are
captured by the non-zero entries in the   $\mathbf{Z_0}$ matrix. In
real scenarios,  the entries in  this matrix are often  extremely
skewed, and the small number of non-zero entries very obviously
expose the outliers.  The decomposition of the matrix into different
component is pictorially illustrated  in Figure
\ref{fig:outlieroverview}.

In order to determine the best low rank factorization, one must try
to optimize the  aggregate  values of the residuals in  the matrix.
This can of course be done in a variety of ways, depending upon the
goals of the underlying factorization process.  We model the
determination of the matrices $\mathbf{W}$,$\mathbf{H}$, and
$\mathbf{Z}$, as the following optimization problem:
\begin{equation}\label{l12norm}
(\mathbf{W}_0,\mathbf{H}_0;\mathbf{Z}_0)=\argmin_{\mathbf{W}\ge0,\mathbf{H}\ge0; \mathbf{Z}}  \frac{1}{2}\norm{\mathbf{A}-\mathbf{W}\mathbf{H}-\mathbf{Z}}_F^2 + \alpha\norm{\mathbf{Z}}_{1,2}
\end{equation}

The specific location of outliers in each column does not have a
closed form solution, since the $\ell_{1,2}$-norm penalty is applied
to $\mathbf{Z}$.   The   logic for applying the $\ell_{1,2}$-norm in
the context of the outlier detection problem is as follows.  Each
entry in the $\mathbf{Z}$ corresponds to a term in a document,
whereas we are interested in the outlier behavior of entire
document. This aggregate outlier behavior of the document $x$ can
be modeled with the $\ell_2$ norm score of a particular column
$\mathbf{z}_x$. In a real scenario,  if a large segment  of a
document $x$ is not representable as the linear combination of the
$r$ topics through $\mathbf{L_0}$, the corresponding column
$\mathbf{z}_x$ in the matrix $\mathbf{Z}$ will be compensated by
having more entries in its column.  In other words, we will  have a
higher $\ell_2$ value for the corresponding column $\mathbf{z}_x$,
and this corresponds to a higher outlier score.
 Furthermore, the  $\ell_{1,2}$-norm penalty on $\mathbf{Z}$
defines the sum of the $\ell_2$ norm outlier scores over all the
documents.  Therefore,  the optimization problem essentially tries
to find the best model,  an important component of which is to
minimize the sum of the outlier scores over all documents.  While a
variety of different  (and more commonly used) penalties such as the
Frobenius norm are available for matrix factorization models, we
have chosen the $\ell_{1,2}$-norm penalty because of its intuitive
significance in the context of the outlier detection problem, and
its tendency to create skewed outlier scores across the columns of
the matrix. As we will see in the next section, this comes at the
expense of a formulation which is more difficult to solve
algorithmically.

For high dimensional data, sparse coefficients are desirable for
obtaining an interpretable low rank matrix $\mathbf{W}\mathbf{H}$.
For this purpose, we add the $\ell_1$-penalty on $\mathbf{H}$:
\begin{equation}\label{outlier}
\min_{\mathbf{W}\ge0,\mathbf{H}\ge0; \mathbf{Z}}  \frac{1}{2}\norm{\mathbf{A}-\mathbf{W}\mathbf{H}-\mathbf{Z}}_F^2 + \alpha\norm{\mathbf{Z}}_{1,2} + \beta\norm{\mathbf{H}}_{1}
\end{equation}
 The constant $\alpha$ defines the weight
for the outlier matrix $\mathbf{Z}$ over the recovery of the low
rank space $\mathbf{L}$ and the sparsity term. In the case of
outlier detection in text documents, we give more weight for the
outlier matrix over the low rank representation $\mathbf{L}$. This
problem does not have a closed form solution, and
 therefore we cannot  directly recover the low rank
matrix $\mathbf{W}\mathbf{H}$ in closed form. However, we can
recover the column space. Without non-negativity constraints, this
property is also known as the rotational invariant property
\cite{ding06,xu12}. This particular formulation of the matrix
factorization model is a bit different from the commonly used
formulations, and off-the-shelf solutions do not directly exist for
this scenario. Therefore, in a later section, we will carefully
design an algorithm with the use of block coordinate descent for
this problem.

\begin{figure}
\includegraphics[scale=0.4]{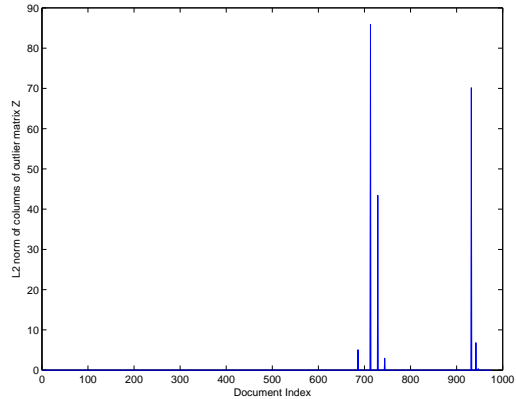}
\caption{$\ell_2$ norm of columns of $\mathbf{Z}$ outlier matrix}
\label{fig:outlierz}
\end{figure}

In order to understand the modeling of the outliers better, we
present the readers with a toy example from a real world data set,
to show how  skewed the typical values of the corresponding column
$\mathbf{z}(x)$ may be in real scenarios.  In this case, we used the
{\em BBC} dataset\footnote{\url{http://mlg.ucd.ie/datasets/bbc.html}}.
\ramki{This dataset consists of documents from BBC news website corresponding
to stories in area business, entertainment, politics, sport, tech from 2004-2005 . 
%We ran our algorithm \algo explained in the next Section
%\ref{sec:algorithm} to find the matrices $\mathbf{W,H}$ and $\mathbf{Z}$.
We took all the documents from business and politics  and 50
documents from tech labeled as outliers. We randomly permuted the columns to 
shuffle the outliers in the matrix to avoid any spatial bias}. 
We computed the $\mathbf{Z}$ matrix
and   generated the $\ell_2$ scores of the columns of outlier matrix
$\mathbf{Z}$. Figure \ref{fig:outlierz} shows the outlier($\ell_2$)
scores of the documents. The $X$-axis illustrates the index of the
document, and the $Y$-axis illustrates the outlier score. It is
evident that  the scores for some  columns are so close to zero,
that they cannot even be seen on the diagram drawn to scale. These
columns also happened to be the non-outlier/regular documents of the collection.
Such documents $\mathbf{a}_x \in \mathbb{R}^m$ correspond to the low
rank space, and  are approximately representable as  a product of
the basis  matrix $\mathbf{W}$ with the corresponding column vector
of coefficients $\mathbf{h}_x \in \mathbb{R}^r$ drawn from
$\mathbf{H}$. However, the documents that are not representable in
such a low rank space have a large outlier score. From the
distribution of the outlier score, we can also observe that the
scores of outlier documents against non-outliers are clearly
separable, by using a simple statistical mean and standard deviation
analysis. Therefore, while we use the scores to rank the documents
in terms of their outlier behavior, the skew in the entries ensures
that it is often easy to choose a cut-off in order to distinguish
the outliers from the non-outliers.

In the following sections, we will analyze the property and
performance of this model \eqref{outlier} for outlier detection
problems.

\section{Algorithmic Solution}
\label{sec:algorithm}

%As discussed in Section \ref{sec:motivation}, our technique is based on NMF.  
As discussed earlier our technique is based on NMF, and
this particular formulation \eqref{outlier}, which is
suited to outlier analysis, is relatively uncommon, and  does not
have a closed form solution. In order to address this issue we use a
Block Coordinate Descent (BCD) framework and its application to
solve the optimization problem \eqref{outlier}. The BCD framework is
a popular choice not only because of the ease in implementation,
but also because it is scalable. First, we will lay the foundation
for the basic BCD technique, as it generally applies to non-linear
optimization problems. We will then relate it to our non-negative
matrix factorization problem, and explain our algorithm \algofull
(\algo) in detail.

\subsection{Block coordinate Descent}

In this section, we will see relevant foundation for using  this
framework.  Consider a constrained non-linear optimization problem
as follows:
\begin{gather}
\min f(x)\:\mbox{ subject to }\:x \in\mathcal{X},\label{eq:general_nonlinear}
\end{gather}
Here,  $\mathcal{X}$ is a closed convex subset of $\mathbb{R}^{n}$.
An important assumption to be exploited in the BCD method is that
the set $\mathcal{X}$ is represented by a Cartesian product:
\begin{equation}
\mathcal{X}=\mathcal{X}_{1}\times\cdots\times\mathcal{X}_{m},\label{eq:bcd-cartesian-product}
\end{equation}
where $\mathcal{X}_{j}$, $j=1,\cdots,m$, is a closed convex subset
of $\mathbb{R}^{N_{j}}$, satisfying $n=\sum_{j=1}^{m}N_{j}$.
Accordingly, the vector $\mathbf{x}$ is partitioned as
$\mathbf{x}=(\mathbf{x}_{1},\cdots,\mathbf{x}_{m})$ so that
$\mathbf{x}_{j}\in\mathcal{X}_{j}$ for $j=1,\cdots,m$. The BCD
method solves for $\mathbf{x}_{j}$ by fixing all other subvectors of
$\mathbf{x}$ in a cyclic manner. That is, if
$\mathbf{x}^{(i)}=(\mathbf{x}_{1}^{(i)},\cdots,\mathbf{x}_{m}^{(i)})$
is given as the current iterate at the $i^{th}$ step, the algorithm
generates the next iterate
$\mathbf{x}^{(i+1)}=(\mathbf{x}_{1}^{(i+1)},\cdots,\mathbf{x}_{m}^{(i+1)})$
block by block, according to the solution of the following
subproblem:
\begin{equation}
\mathbf{x}_{j}^{(k+1)}\leftarrow\underset{\mathbf{\xi}\in\mathcal{X}_{j}}{\text{argmin}}
f(\mathbf{x}_{1}^{(k+1)},\cdots,\mathbf{x}_{j-1}^{(k+1)},\mathbf{\xi},\mathbf{x}_{j+1}^{(k)},\cdots,\mathbf{x}_{m}^{(k)}).\label{eq:bcd-method}
\end{equation}
Also known as a \textit{non-linear Gauss-Seidel}
method~\cite{Bertsekas1999}, this algorithm updates one block each
time, always using the most recently updated values of other blocks
$\mathbf{x}_{\tilde{j}},\tilde{j}\ne j$. This is important since it
ensures that after each update, the objective function value does
not increase. For a sequence $\left\lbrace
\mathbf{x}^{(i)}\right\rbrace $ where each $\mathbf{x}^{(i)}$ is
generated by the BCD method, the following property holds.
\begin{thm}
\label{thm:bcd}
Suppose $f$ is continuously differentiable in $\mathcal{X}=\mathcal{X}_{1}\times\dots\times\mathcal{X}_{m}$,
where $\mathcal{X}_{j}$, $j=1,\cdots,m$, are closed convex sets.
Furthermore, suppose that for all $j$ and $i$, the minimum of
\[
\min_{\mathbf{\mathbf{\xi}}\in\mathcal{X}_{j}}f(\mathbf{x}_{1}^{(k+1)},\cdots,\mathbf{x}_{j-1}^{(k+1)},\mathbf{\xi},\mathbf{x}_{j+1}^{(k)},\cdots,\mathbf{x}_{m}^{(k)})
\]
is uniquely attained. Let $\left\lbrace
\mathbf{x}^{(i)}\right\rbrace $ be the sequence generated by the
block coordinate descent method as in Eq.~\eqref{eq:bcd-method}.
Then, every limit point of $\left\lbrace
\mathbf{x}^{(i)}\right\rbrace $ is a stationary point. The
uniqueness of the minimum is not required for the case when  $m=2$
\cite{Grippo2000}.
\end{thm}

The proof of this theorem for an arbitrary number of blocks is shown
in Bertsekas~\cite{Bertsekas1999}.
For a non-convex optimization problem, most algorithms only guarantee
the stationarity of a limit point \cite{Lin2007}.

When applying the BCD method to a constrained non-linear programming
problem, it is critical to wisely choose a partition of
$\mathcal{X}$, whose Cartesian product constitutes $\mathcal{X}$. An
important criterion is whether the sub-problems in
Eq.~\eqref{eq:bcd-method} are efficiently solvable. For example, if
the solutions of sub-problems appear in a closed form, each update
can be computed fast. In addition, it is worth checking how the
solutions of sub-problems depend on each other. The BCD method
requires that the most recent values be used for each sub-problem in
Eq.~\eqref{eq:bcd-method}. When the solutions of sub-problems depend
on each other, they have to be computed sequentially to make use of
the most recent values. If solutions for some blocks are independent
of each other,  they can be computed simultaneously. We discuss how
different choices of partitions lead to different NMF algorithms.
The  partitioning can be achieved in several ways, by using either
matrix blocks, vector blocks or scalar blocks.

\subsubsection{BCD with Two Matrix Blocks - ANLS Method}
The most natural partitioning of the variables is to have two big
blocks, $\mathbf{W}$ and $\mathbf{H}$. In this case, following the
BCD method in Eq.~\eqref{eq:bcd-method}, we take turns solving the
following:
%\begin{equation}
%\left\{
%\begin{array}
%\mathbf{W}^{(k+1)}\leftarrow\argmin_{\mathbf{W}\ge0}f(\mathbf{W},\mathbf{H}^{(k)})\\ 
%\mathbf{H}^{(k+1)}\leftarrow\argmin_{\mathbf{H}\ge0}f(\mathbf{W}^{(k+1)},\mathbf{H}).
%\end{array}
%\right.
%\end{equation}

\begin{equation}
\left\{
  \begin{array}{ll}
  \mathbf{W}^{(k+1)}\leftarrow\argmin_{\mathbf{W}\ge0}f(\mathbf{W},\mathbf{H}^{(k)})\\ 
\mathbf{H}^{(k+1)}\leftarrow\argmin_{\mathbf{H}\ge0}f(\mathbf{W}^{(k+1)},\mathbf{H}).  
\end{array}
\right.\label{bcd}
\end{equation}

Since the sub-problems are non-negativity constrained least squares (NLS) problems, the
two-block BCD method has been called the alternating non-negative least
square (ANLS) framework \cite{Lin2007,Kim2008,Kim2011}.

\subsubsection{BCD with 2k Vector Blocks - HALS/RRI Method}
We partition the unknowns into 2k blocks in which each block
is a column/row of $\mathbf{W}$ or $\mathbf{H}$. In this case,
it is easier to consider the objective function in the following
form:
\begin{equation}
f(\mathbf{w}_{1},\cdots,\mathbf{w}_{r},\mathbf{h}_{1}^\intercal,\cdots,\mathbf{h}_{r}^\intercal)=\|\mathbf{A}-\sum_{j=1}^{r}\mathbf{w}_{j}\mathbf{h}_{j}^{T}\|_{F}^{2},\label{eq:nmf_columns_obj}
\end{equation}
where
$\mathbf{W}=[\mathbf{w}_{1},\cdots\mathbf{w}_{r}]\in\mathbb{R}_{+}^{m\times
r}$ and
$\mathbf{H}=[\mathbf{h}_{1},\cdots,\mathbf{h}_{r}]^\intercal\in\mathbb{R}_{+}^{r\times
n}$. The form in Eq.~\eqref{eq:nmf_columns_obj} represents the fact
that  $\mathbf{A}$ can be  approximated by the sum of $r$ rank-one
matrices.

Following the BCD scheme, we can minimize $f$ by iteratively solving
the following:
\[
\mathbf{w}_{i}\leftarrow\argmin_{\mathbf{w}_{i}\ge0}f(\mathbf{w}_{1},\cdots,\mathbf{w}_{r},\mathbf{h}_{1}^\intercal,\cdots,\mathbf{h}_{r}^\intercal)
\]
 for $i=1,\cdots,r$, and
\[
\mathbf{h}_{i}^\intercal\leftarrow\argmin_{\mathbf{h}_{i}^\intercal\ge0}f(\mathbf{w}_{1},\cdots,\mathbf{w}_{r},\mathbf{h}_{1}^\intercal,\cdots,\mathbf{h}_{r}^\intercal)
\]
 for $i=1,\cdots,r$.

% \begin{figure}
% \scalebox{0.42}{\includegraphics{figs/2kblock.eps}}
% \caption{2k-Block Block Coordinate Descent(BCD)}\label{fig:2kblock}
% \end{figure}

The  2K-block BCD algorithm has been studied as Hierarchical
Alternating Least Squares (HALS) proposed by Cichocki et al.
\cite{Cichocki2007,Cichocki2009} and independently by Ho et al.
\cite{Ho2008} as rank-one residue iteration (RRI).

\subsubsection{BCD with k(n + m) Scalar Blocks}
 We can also partition the variables with the smallest  $k(n+m)$ element blocks of scalars, where every element of $\mathbf{W}$ and $\mathbf{H}$ is considered as a block in the context of \ref{thm:bcd}.   To this end, it
helps to write the objective function as a quadratic function of scalar
$w_{ij}$ or $h_{ij}$ assuming all other elements in $\mathbf{W}$
and $\mathbf{H}$ are fixed: \begin{subequations} \label{eq:nmf_element_objective}
\begin{gather}
f(w_{ij})  =  \|(\mathbf{a}_{i}^\intercal-\sum_{\tilde{k}\ne j}w_{i\tilde{k}}\mathbf{q}_{\tilde{k}}^\intercal)-w_{ij}\mathbf{h}_{j}^\intercal\|_{2}^{2}+\mbox{const},\\
f(h_{ij})  =  \|(\mathbf{a}_{j}-\sum_{\tilde{k}\ne i}\mathbf{w}_{\tilde{k}} h_{\tilde{k}j})-\mathbf{w}_{ i}h_{ij}\|_{2}^{2}+\mbox{const},
\end{gather}
\end{subequations} where $\mathbf{a}_{i}^\intercal$ and $\mathbf{a}_{j}$
denote the $i^{th}$ row and the $j^{th}$ column of $\mathbf{A}$,
respectively.

%\begin{figure}
%\scalebox{0.35}{\includegraphics{figs/scalarblock.eps}}
%\caption{Scalar Block BCD}\label{fig:scalarblock}
%\end{figure}

In this paper for solving the optimization problem \eqref{outlier},
we partition the matrices $\mathbf{Z,W,H}$ into vector blocks such
as \linebreak $\mathbf{z_1, \cdots, z_n, w_1, \cdots, w_r, h_1,\cdots, h_r}$. The
reasoning behind this partitioning is explained in the next section.

\subsection{\algofull (\algo)}
In this section, we propose an efficient algorithm for the outlier
detection model \eqref{outlier}.

To determine the $\mathbf{Z,W,H}$ for the aforementioned
optimization problem \eqref{outlier}, we use the  block coordinate
descent method. In other words, by fixing $\mathbf{W,H}$, we
determine the optimal $\mathbf{Z}$ as vector blocks
$\mathbf{z_1,\cdots,z_n}$ and vice versa. Due to $\ell_{1,2}$-norm,
this optimization corresponds to the two block non-smooth BCD
framework.

\begin{equation}
\begin{aligned}
\label{am}
\mathbf{Z}^{(k+1)} & \leftarrow \underset{\mathbf{Z}}{\argmin} \frac{1}{2}\norm{\mathbf{A}-\mathbf{Z}-\mathbf{W}^{(k)}\mathbf{H}^{(k)}}_F^2 \\
                    &\qquad \qquad \qquad \qquad \qquad \qquad \quad     + \alpha \norm{\mathbf{Z}}_{1,2}    \\
(\mathbf{W}^{(k+1)},\mathbf{H}^{(k+1)}) & \leftarrow \underset{\mathbf{W\geq0}, \mathbf{H\geq0}}{\argmin} \frac{1}{2}\norm{\mathbf{A}-\mathbf{WH}-\mathbf{Z}^{(k+1)}}  \\
                                        & \qquad \qquad \qquad \qquad \qquad \qquad \quad + \beta \norm{\mathbf{H}}_1 &    \\
\end{aligned}
\end{equation}
%We have a closed form solution for outlier variable $\mathbf{Z}$. See \ref{thm:updatez}. And, for low rank matrix $\mathbf{W}\mathbf{H}$, we need to solve highly non-convex problems.
Regarding ${\bf Z}=[{\bf z}_1,...,{\bf z}_n]$, the minimization
problem in \eqref{am} has a  separable structure:
$$
\mathbf{Z}^{(k+1)} =  \underset{\mathbf{Z}}{\argmin} \sum_i \frac{1}{2}\norm{\mathbf{\bar{a}}_i -\mathbf{z}_i}_2^2+ \alpha \norm{\mathbf{z}_i}_{2}
$$
where $\mathbf{\bar{a}}_i = {\bf a}_i - ({\bf W}^{(k)}{\bf
H}^{(k)})_i$. Therefore, we only need to define a solution with
respect to one variable ${\bf z}_i$. Thus, we partition the matrix
$\mathbf{Z}$ into vector blocks $\mathbf{z}_i$ and construct
$\mathbf{Z}$ as a set of vectors $\mathbf{z}_i$. Also, the blocks
$\mathbf{z}_i$ is independent of $\mathbf{z}_j, \forall i \neq j$.
That is, the closed form solution of $\mathbf{z}_i$ is dependent
only on $\mathbf{\bar{a}}_i$. When all other blocks of
$\mathbf{w_1,\cdots,w_r,h_1,\cdots,h_r}$, are fixed, every vector
$\mathbf{z}_i \in \mathbf{Z}$, can be solved to optimal in parallel.
Thus, we adhere to BCD framework of solving the vector blocks of
$\mathbf{z}_i$, to optimal, when all the other blocks are fixed.

%Even individually solving, $\mathbf{Z,H}$ are non-convex in the above problem. Hence, let us define a (n+2k) vector block BCD on the above problem.

\begin{thm}\label{thm:updatez}
The solution of the following minimization problem
$$
\mathbf{z}_i^* = \argmin_{\mathbf{z}_i} f({\bf z}_i) = \frac{\gamma}{2}\norm{{\bf z}_i-{\bf a}_i}_2^2 + \alpha \norm{{\bf z}_i}_2
$$
is the generalized shrinkage operator:
$$
{\bf z}_i^* = \mbox{shrink}({\bf a}_i , \frac{\alpha}{\gamma})
$$
where generalized shrinkage operator is defined as:
$$
\mbox{shrink}({\bf a}_i,C) = \mbox{max}(\norm{{\bf a}_i}_2 -C,0)\frac{{\bf a}_i}{\norm{{\bf a}_i}_2}
$$
\end{thm}
\vspace{-0.8in}
\begin{proof}
$$
\frac{\partial f({\bf z}_i)}{\partial {\bf z}_i} = \gamma ({\bf z}_i - {\bf a}_i) + \alpha \frac{{\bf z}_i}{\norm{{\bf z}_i}}
$$
When $\norm{{\bf a}_i}_2 \le \frac{\alpha}{\gamma}$,
$$
f({\bf z}_i) \ge \frac{\gamma}{2}(\norm{{\bf z}_i}^2_2 + \norm{{\bf a}_i}^2_2)  + (\alpha-\gamma\norm{{\bf a}_i}_2)\norm{{\bf z}_i}_2
$$
Therefore we have:
$$
\argmin_{{\bf z}_i} f({\bf z}_i) =0.
$$

When $\norm{{\bf a}_i}_2 \ge \frac{\alpha}{\gamma}$,
let ${\bf z}_i = c{\bf a}_i$ then
$$
\frac{\partial f({\bf z}_i)}{\partial {\bf z}_i} = \gamma ({\bf z}_i - {\bf a}_i) + \alpha \frac{{\bf z}_i}{\norm{{\bf z}_i}_2} = [ \gamma( c - 1 ) + \frac{\alpha}{\norm{{\bf a}_i}}_2 ]{\bf a}_i=0
$$
where
$$
c =  1 - \frac{\alpha}{\gamma}\frac{1}{\norm{{\bf a}_i}_2}.
$$
Therefore, we get
$$
{\bf z}_i =  (\norm{{\bf a}_i}_2 - \frac{\alpha}{\gamma})\frac{{\bf a}_i}{\norm{{\bf a}_i}_2}
$$
Now, utilizing the generalized shrinkage operator as defined in \cite{esser2010}\cite{wang2008},
$$
{\bf z}_i^* = \mbox{shrink}({\bf a}_i,C) = \mbox{max}(\norm{{\bf a}_i}_2 -C,0)\frac{{\bf a}_i}{\norm{{\bf a}_i}_2}
$$
where $C=\alpha/\gamma$.
\end{proof}

%We will find $\mathbf{Z}=(\mathbf{z}_1,\mathbf{z}_2,\cdots,\mathbf{z}_n)$ as separate $n$ vector blocks. Given $\mathbf{A,W,H}$, the optimal value of $\mathbf{z}_i= max(\mathbf{||d||_i}-\alpha,0) \times \frac{\mathbf{d_i}}{||\mathbf{d}_i||}$, where $\mathbf{d}_i$ is the $i$-th column of matrix $\mathbf{D=A-WH}$.
%\begin{proof}
%We will find the $\mathbf{Z}$ as separate $n$ vector blocks. Then, the optimal value of $\mathbf{Z=(z_1,z_2,\cdots z_n)}$ in equation \eqref{bcdoutlier} can be rewritten as
%\begin{equation}
%\begin{aligned}
%\label{bcdupdatez}
%f_\mathbf{z}=\underset{\mathbf{Z}}{argmin}&\sum_{i=1}^n\frac{1}{2}||\mathbf{d}_i-\mathbf{z}_i||^2+\alpha ||\mathbf{z}_i|| \\
%\underline{if \; ||\mathbf{d}_i||_2 > \alpha} \\
%\frac{\partial f_\mathbf{z}}{\partial \mathbf{z}_i}=&
%\mathbf{d}_i-\mathbf{z}_i + \lambda \frac{\mathbf{z}_i}{||\mathbf{z}_i||}   \\
%= & \alpha (\frac{\mathbf{d}_i}{||\mathbf{d}_i||} - \frac{\mathbf{z}_i}{||\mathbf{z}_i||}) \\
%\underline{if \; ||\mathbf{d}_i||_2 < \alpha}\\
%& \frac{1}{2}||\mathbf{d}_i-\mathbf{z}_i||^2+\alpha ||\mathbf{z}_i|| \\
%\geq & \frac{1}{2} ||\mathbf{d}_i|| + \frac{1}{2} ||\mathbf{z}_i|| - ||\mathbf{d}_i|| ||\mathbf{z}_i|| + \alpha ||\mathbf{z}_i|| \\
%\geq & 0 \\
%\end{aligned}
%\end{equation}
%From the above, we can conclude that $\mathbf{z}_i= max(\mathbf{||d||_i}-\alpha,0) \times \frac{\mathbf{d_i}}{||\mathbf{d}_i||}$.
%\end{proof}

%It is interesting to observe here that every $\mathbf{z}_i$ is independent of $\mathbf{z}_j, \forall j \neq i$. Hence the update of $\mathbf{Z}$ can be done in parallel.

Now, we need to solve the following NMF model with sparsity constraints on ${\bf H}$:
$$
({\bf W}^{(k+1)},{\bf H}^{(k+1)}) = \argmin_{{\bf W} \ge 0, {\bf H} \ge 0}  \norm{\bar{\bf A} - {\bf W}{\bf H}}_F^2 + \beta \norm{\bf H}_1
$$
where $\bar{\bf A} = {\bf A} - {\bf Z}^{(k+1)}$. Let
\begin{equation}\label{hals}
{\cal F}({\bf w}_1,...,{\bf w}_r;{\bf h}_1,...,{\bf h}_r) = \norm{\bar{\bf A} - \sum_{i=1}^r {\bf w}_i{\bf h}_i}_F^2 + g({\bf h}_1,...,{\bf h}_r).
\end{equation}
where ${\bf W} = [{\bf w}_1,{\bf w}_2,\dots,{\bf w}_r]$ and ${\bf H} = [{\bf h}_1, {\bf h}_2, \dots, {\bf h}_r]^T$. For any $j \in \{1,...,r\}$, (\ref{hals}) can be rewritten as
\begin{equation}\label{hals-1}
\norm{\bar{{\bf A}}-\sum_{i=1}^r{\bf w}_i{\bf h}_i^T}^2_F=\norm{\bar{\bf A}-\sum_{i=1,i\not=j}^r {\bf w}_i{\bf h}_i^T-{\bf w}_j{\bf h}_j^T}^2_F.
\end{equation}
The following is the framework of the  block coordinate descent
method with a separable regularizer such as the  Frobenius norm. We
iteratively minimize ${\cal F}({\bf W},{\bf H})$ with respect to
each column of ${\bf W}$ and ${\bf H}$ :
\begin{equation}
\left\{
  \begin{array}{ll}
  {\rm for}\ j=1 \dots r \\[5pt]
  {\bf h}_j^{(k+1)} = \underset{{\bf h}_j \ge 0}{\rm argmin} \frac{\alpha}{2}\norm{{\bf w}_j^{(k)}{\bf h}_j^T-(\bar{\bf A}-\tilde{\bf W}^{(k)}_j)}^2_F\\\hspace{3cm} + g({\bf h}_1^{(k+1)},...,{\bf h}_j,...,{\bf h}_r^{(k)}) \\[5pt]
  {\rm end}\\[5pt]
    {\rm for}\ j=1 \dots r \\[5pt]
  \hspace{0.25cm}{\bf w}_j^{(k+1)} = \underset{{\bf w}_j \ge 0}{\rm argmin} \norm{{\bf w}_j ({\bf h}_j^{(k+1)})^T -(\bar{\bf A}-\tilde{\bf H}^{(k+1)}_j)}^2_F \\[5pt]
  {\rm end}\\
\end{array}
\right.\label{halsalter}
\end{equation}
where
$$
\tilde{\bf W}^{(k)}_j = \sum_{i=1}^{j-1}{\bf w}_i^{(k)}({\bf h}_i^{(k+1)})^T + \sum_{i=j+1}^r{\bf w}_i^{(k)}({\bf h}_i^{(k)})^T,
$$
and
$$
\tilde{\bf H}^{(k+1)}_j = \sum_{i=1}^{j-1}{\bf w}_i^{(k+1)}({\bf h}_i^{(k+1)})^T + \sum_{i=j+1}^r{\bf w}_i^{(k)}({\bf h}_i^{(k+1)})^T.
$$

According to \ref{thm:updatez}, the solution of $\mathbf{z}_i$ is independent of
$\mathbf{z}_j, \forall i \neq j $,  and it enables us to solve the
solution in parallel. This is very useful when computing for very
large input matrices. Similarly, the vector blocks of $\mathbf{W, H}$
can also be updated in parallel. Now, we have all the building
blocks for the {\algofull}  algorithm. We will be using
\ref{thm:updatez} and the update for $\mathbf{W,H}$ from
\eqref{halsalter}. The Algorithm \ref{alg:tomf}, gives the outline
of the \algo and its complete implementation can be obtained 
from \url{https://github.com/ramkikannan/outliernmf} to try with
any real world text dataset. 

\begin{algorithm}
  \SetKwInOut{Input}{input}\SetKwInOut{Output}{output}
  \Input{Matrix $\mathbf{A}\in\mathbb{R}_+^{m \times n}$,reduced rank $r$, $\alpha$, $\beta$}
  \Output{Matrix $\mathbf{W}\in\mathbb{R}_+^{m\times r}$,$\mathbf{H}\in\mathbb{R}_+^{r \times n}$,$\mathbf{Z} \in \mathbb{R}^{m \times n}$}
  \BlankLine
    \tcp{Rand initialization of \textbf{W}, \textbf{H}, \textbf{Z}}
    %\tcp{ For other types refer to Section \ref{sec:initialization}}
  \textit{Initialize \textbf{W}, \textbf{H}, \textbf{Z} as a nonnegative random matrix} \;
  \While{stopping criteria $\mathfrak{C}_1$ not met} {
  \tcp{Compute Z for the given $\mathbf{A,W,H},\alpha,\beta$ based on \ref{thm:updatez}} 
  \For{$i \leftarrow 1$ \KwTo $n$}{ \label{alg:loopz}
    $\mathbf{z_i} \leftarrow max(\norm{{\bf a}_i}_2 - \frac{\alpha}{\gamma},0)\frac{{\bf a}_i}{\norm{{\bf a}_i}_2}$
  }
  \While{stopping criteria $\mathfrak{C}_2$ not met}{ \label{alg:innerbcd}
      \For{$j \leftarrow 1$ \KwTo $r$}{
        $ {\bf h}_j^{(k+1)} = \underset{{\bf h}_j \ge 0}{\rm argmin} \frac{\alpha}{2}\norm{{\bf w}_j^{(k)}{\bf h}_j^T-(\bar{\bf A}-\tilde{\bf W}^{(k)}_j)}^2_F+ g({\bf h}_1^{(k+1)},\cdots,{\bf h}_j,\cdots,{\bf h}_r^{(k)})$\;
    }
    \For{$j \leftarrow 1$ \KwTo $r$}{
        ${\bf w}_j^{(k+1)} = \underset{{\bf w}_j \ge 0}{\rm argmin} \norm{{\bf w}_j ({\bf h}_j^{(k+1)})^T -(\bar{\bf A}-\tilde{\bf H}^{(k+1)}_j)}^2_F$\;
    }
  }
}
\caption{\algofull\ (\algo)}\label{alg:tomf}
\end{algorithm}

\section{Experimental Results}
\label{sec:experimentation}

In this section, we present the experiments on text outlier analysis
using matrix factorization. We used both real and synthetic data
sets to test our algorithm. The  real data sets correspond to the
well known {\em RCV20}, {\em Reuters} and {\em Wiki People} data, whereas the
synthetic data set was created using a well known market basket
generator described later.  It should be pointed out that these data
sets were not originally designed for outlier analysis, and they
have no ground truth information available.  Therefore, some
additional pre-processing needed to be applied to the real data
sets, in order to isolate ground truth classes, and use them
effectively for the outlier analysis problem. In this section, we
will describe the data sets, their preparation, the performance
criteria and the results obtained by our algorithm. At the end of
this section, we will also present a discussion that provides
interesting insights about  the effectiveness of algorithm \algo\ .

\subsection{Data Sets}
The experiments were  conducted with both labelled real  and
synthetic data sets.  These are described below:\\
{\bf\em RCV20 Data Set:}  The {\em RCV20 data set}
\footnote{\url{http://qwone.com/~jason/20Newsgroups/}}
 is a collection of approximately 20,000
newsgroup documents, partitioned (nearly) evenly across 20 different
newsgroups. We took all data points from two randomly chosen
classes, which in this case corresponded to the {\em IBM} and {\em
Mac Hardware} classes. In addition, 50 data points were chosen from
one randomly chosen class, which corresponds to the {\em Windows
Operating System (OS)} class. As it turns out, this is a rather hard
problem for  our algorithm because of some level of relationship
between one of the rare classes and the base data. Specifically,
{\em Windows Operating System} and {\em IBM Hardware} are both
computer related subjects, and the former is often used with the
latter. Therefore, some vocabulary is shared between the regular
class and the rare class, and this makes the detection of outlier
harder. We randomly permuted the position of the outliers and
regular data points. \\
{\bf\em Reuters-21578 Data Set:}  The documents in the {\em
Reuters-21578} collection
\footnote{\url{http://archive.ics.uci.edu/ml/datasets/Reuters-21578+Text+Categorization+Collection}}
 appeared on the {\em Reuters} newswire in 1987. It contains  21578 documents in 135
categories. Every document belongs to one or  more categories.
We selected those documents that belong to only one category.
We chose totally 5768 documents that belong to the category {\em earn} and {\em acq}.
The outliers were 100 documents from category {\em interest}. The vocabulary size
of all the documents from these categories put together were 18933.
We randomly permuted the position of the outliers and
regular data points.
\\
\ramki{{\bf\em Wiki People Dataset:} This is a subset of the dataset
collected by Blasiak et.al., \cite{Blasiak13}. The dataset is
constructed by crawling Wikipedia starting from
\url{http://en.wikipedia.org/wiki/Category:Lists_of_politicians} to
a depth of four. Pages describing people were extracted from the
list of all crawled pages. Text from the body paragraphs of the
pages were extracted, and section headings were used as labels for
blocks of text. Text blocks were assumed  to begin with <p> and end
with </p>. Only text in section headings that occurred 10 times or
more was retained. Words were stemmed, stopwords were removed, and
words of length at least 3 and  at most 15 were considered. The
words need to occur at least 4 times in at least 2 documents to be
considered important enough to be  retained. From the collected
data, the sections {\em Career} and {\em Life} were chosen as
non-outlier and whereas the small section  section {\em Death} was
chosen as outlier. The constructed dataset has a vocabulary size of
18834 and total of 9593 documents.  A total of 100 documents that
belong to section {\em Death} were labeled as outlier. }
 \\
{\bf\em  Market Basket Data Generator:} We also wanted to
understand the performance of our algorithm in some large sparse
matrices that is similar to the bag of words matrix. Towards this end, we used the standard {\em
IBM Synthetic Data Generation Code for Associations and Sequential
Patterns} -- market-basket data generator, that is packaged as part
of {\em Illimine}\footnote{\url{http://illimine.cs.uiuc.edu/}}
software. We set the average length of the transaction to be 300 and
number of different items to be 50,000. Note that this generator
uses a random seed, and by changing the seed, it is possible to
completely change the transaction distribution, even if all other
parameters remain the same. We generated 10,000 data points as a
group of four different  sets of 2500 data points with randomly
chosen  seed values. In addition, the rare class contained 250 data
points from a single seed value.  In addition, we randomly permuted
the positions of the outliers and regular data points in the matrix
representation, to avoid any unforeseen bias in the algorithm.

\subsection{Performance Metrics}
The effectiveness was measured in terms of the ROC curve drawn on
the outlier scores. We use the area under the Receiver Operating
Characteristics(ROC) curve -- the defacto metric for evaluation in
outlier analysis. The idea of this curve is to evaluate a {\em
ranking} of outlier scores, by examining the tradeoff between the
true positives and false positives, as the threshold on the outlier
score is varied in a range. By using different thresholds, it is
possible to obtain a relatively larger or smaller number of true
positives with respect to the false positives.

Let $S(t)$ be the set of outliers determined by using a threshold
$t$ on the outlier scores. In this case,  the {\em True Positive
Rate} is graphed against the {\em False Positive Rate}. The true
positive rate $TPR(t)$ is defined in the same way as the metric of
recall is defined in the IR literature. The false positive rate
$FPR(t)$ is the percentage of the falsely reported positives out of
the ground-truth negatives. Therefore, for a data set $D$ with
ground truth positives $G$, these definitions are as follows:
\begin{equation*}
TPR(t)= Recall(t)= 100* \frac{ |S(t) \cap G|}{|G|}
\end{equation*}
\begin{equation*}
FPR(t) =100 * \frac{|S(t) -G|}{|D - G|}
\end{equation*}
 Note that the end points of the ROC curve are
always at $(0, 0)$ and $(100, 100)$, and a random method is expected
to exhibit performance along the diagonal line connecting these
points. The {\em lift} obtained above this diagonal line provides an
idea of the accuracy of the approach. The area under the ROC curve
provides a measure of the accuracy. A random algorithm would have an
area of 0.5 under the ROC curve. The ROC curve was used to provide
detailed insights into the tradeoffs associated with the method,
whereas the area under the ROC curve was used in  order to provide a
summary of the performance of the method.
\subsection{Baseline Algorithms}
The baselines  used by our approach were as follows:\\
{\bf\em Distance-based Algorithm:}  The first algorithm which was
used was the
 $k$-nearest neighbour algorithm, which is a classical distance-based algorithm
 frequently used for outlier detection \cite{knorr,rama}. The
 outliers were ranked based on distances in order to create an ROC
 curve, rather than using a specific threshold as in \cite{knorr}. In addition, we
 gave the $k$-nearest neighbour algorithm an advantage by picking a value of $k$ optimally
 based on area under ROC curve by sweeping $k$ from 1 to 50. Note that such an advantage would
 not be available to the baseline under real scenarios, since the ground-truth
 outliers in the data are unknown, and therefore the ROC curve cannot be optimized.\\
{\bf\em Simplified Low Rank Approximation:} We used a low rank
approximation based on Singular Value Decomposition ($SVD$). For a
 given matrix $\mathbf{A}$, a best $r$-rank approximation $\hat{\mathbf{A}}_r$ is given by
 $\hat{\mathbf{A}}_r = \mathbf{US_rV^\intercal}$, where \linebreak $\mathbf{S_r}=diag(\sigma_1,\cdots,\sigma_r,0,\cdots,0)$.
 That is, the trailing $rank(\mathbf{A})-r$ in the descending ordered singular values are set to $0$.
 It is natural to understand that the outlier documents require
 linear combination of many basis vectors. Thus the $\ell_2$ norm
on the $\sqrt{S_r}V^\intercal$ can be used a score to determine the outliers. In the graphs, we use $SVD$
 as the legend to represent this baseline. For the $SVD$ approach,
 we used the same low rank as our algorithm.

\ramki{ {\bf \em Robust Principal Component Analysis(RPCA)} :
Recently Candes et.al.,\cite{Candes2011}, proposed a new technique
called Robust PCA that is insensitive to noises and outliers. It is
important to note that both PCA and NMF are different forms  of low
rank approximation. Hence, we wanted to leverage the output of RPCA
and recover the outliers. RPCA yields two matrices (1) a low rank
matrix - $\mathbb{L}$ and (2) a sparse matrix $\mathbb{S}$ such that
$\mathbb{A} \approx \mathbb{L+S}$, where $\mathbb{A}$ is the given
input matrix. The main disadvantage of RPCA is its larger  memory
requirements.  Retaining $\mathbb{L, S}$ for large matrices require
significant memory. We used the $\ell_2$ norm on the $\mathbb{S}$ as
an outlier score for every document. In the graphs, we use $RPCA$ as
the legend to represent this baseline.}

\subsection{Effectiveness Results}

\begin{figure*}[htbp]
 \begin{minipage}{0.5\linewidth}
  \centering
  \caption*{ROC}
  \includegraphics[width=2.4in,height=1.8in]{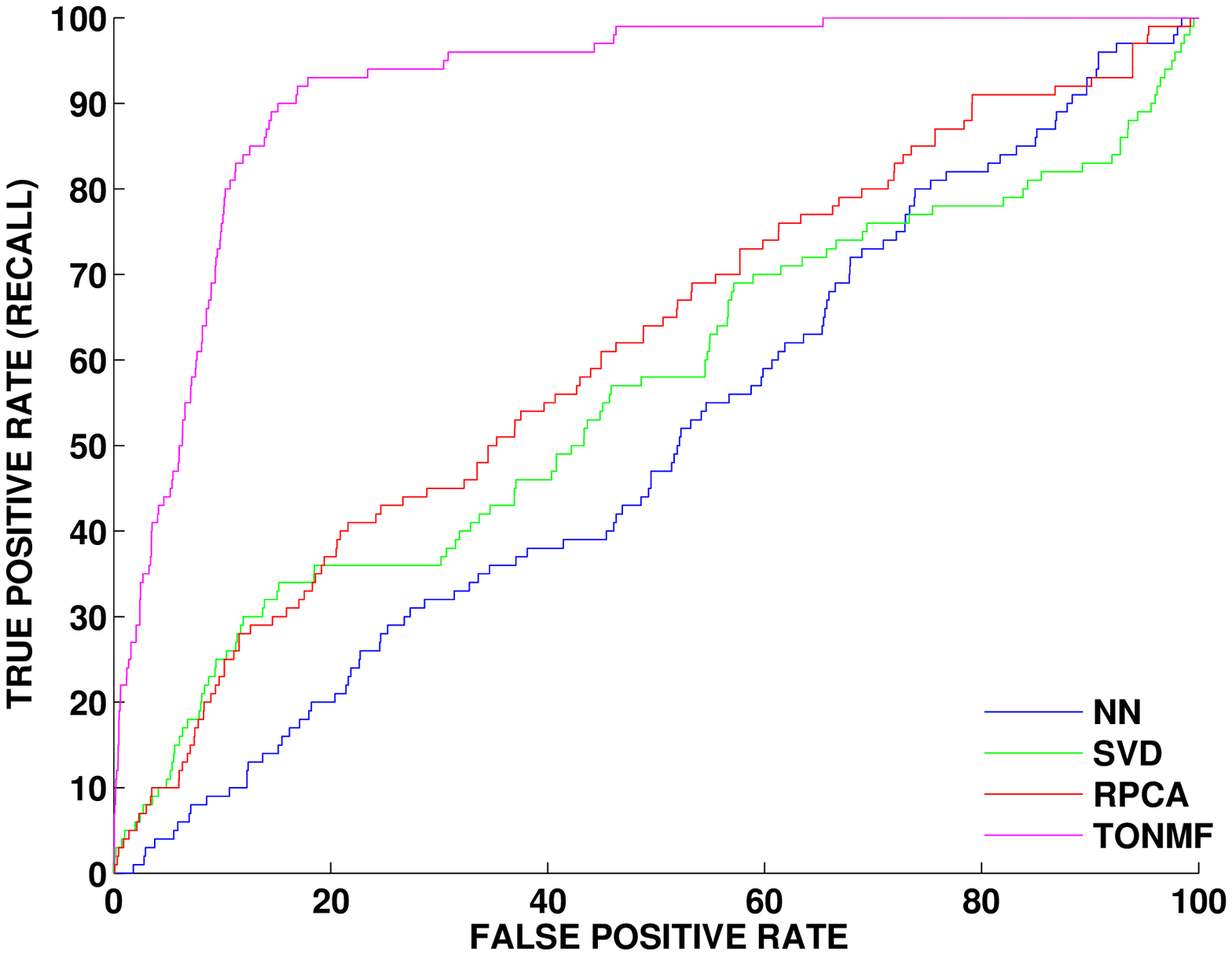}
  \caption{Reuters}
  \label{fig:rocreuters}
 \end{minipage}%
 \begin{minipage}{0.5\linewidth}
  \centering
  \caption*{Parameter Sensitivity}
  \includegraphics[width=2.4in,height=1.8in]{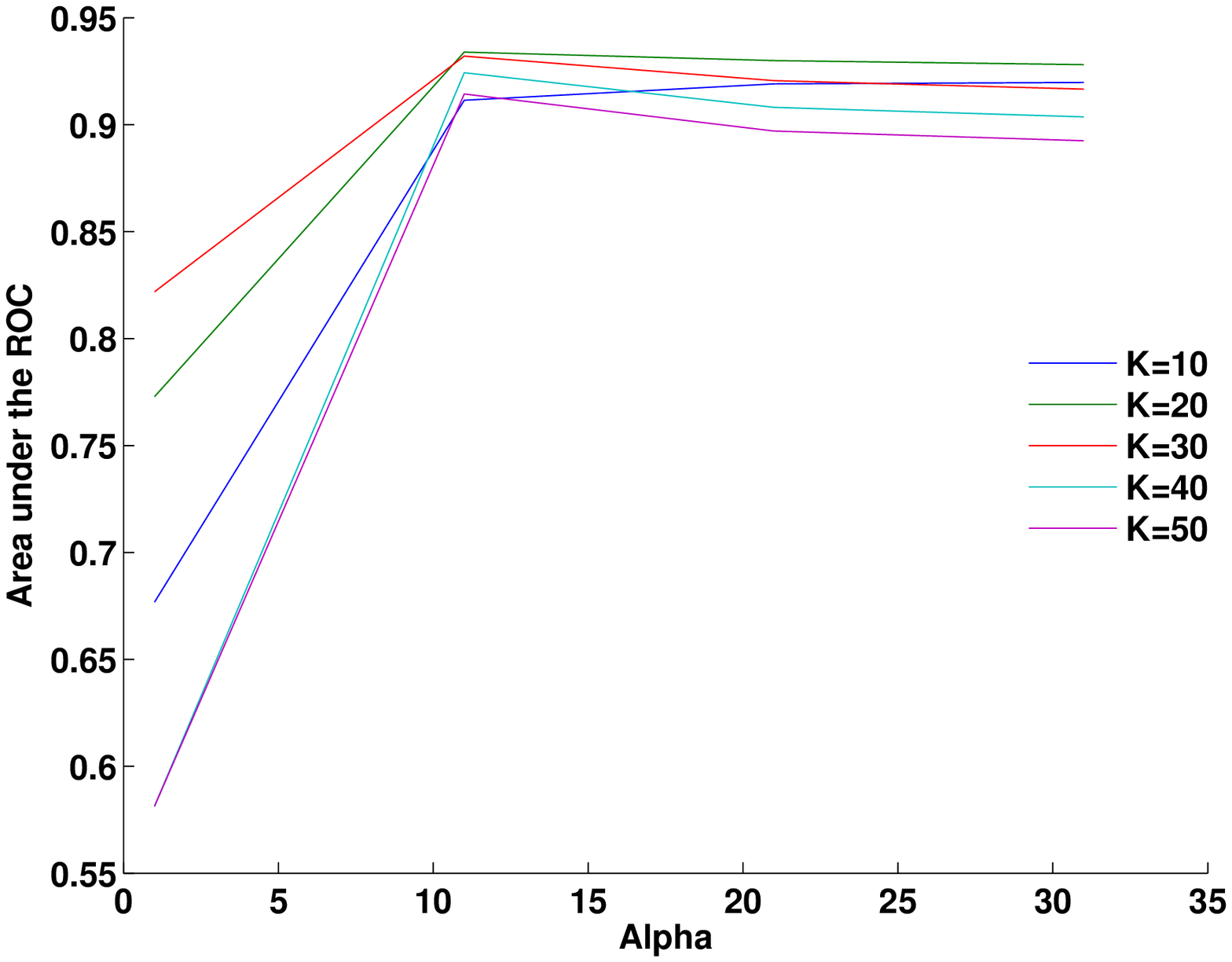}
  \caption{Reuters}
  \label{fig:alphakreuters}
 \end{minipage}
 \begin{minipage}{0.5\linewidth}
  \centering
  \caption*{ROC}
  \includegraphics[width=2.4in,height=1.8in]{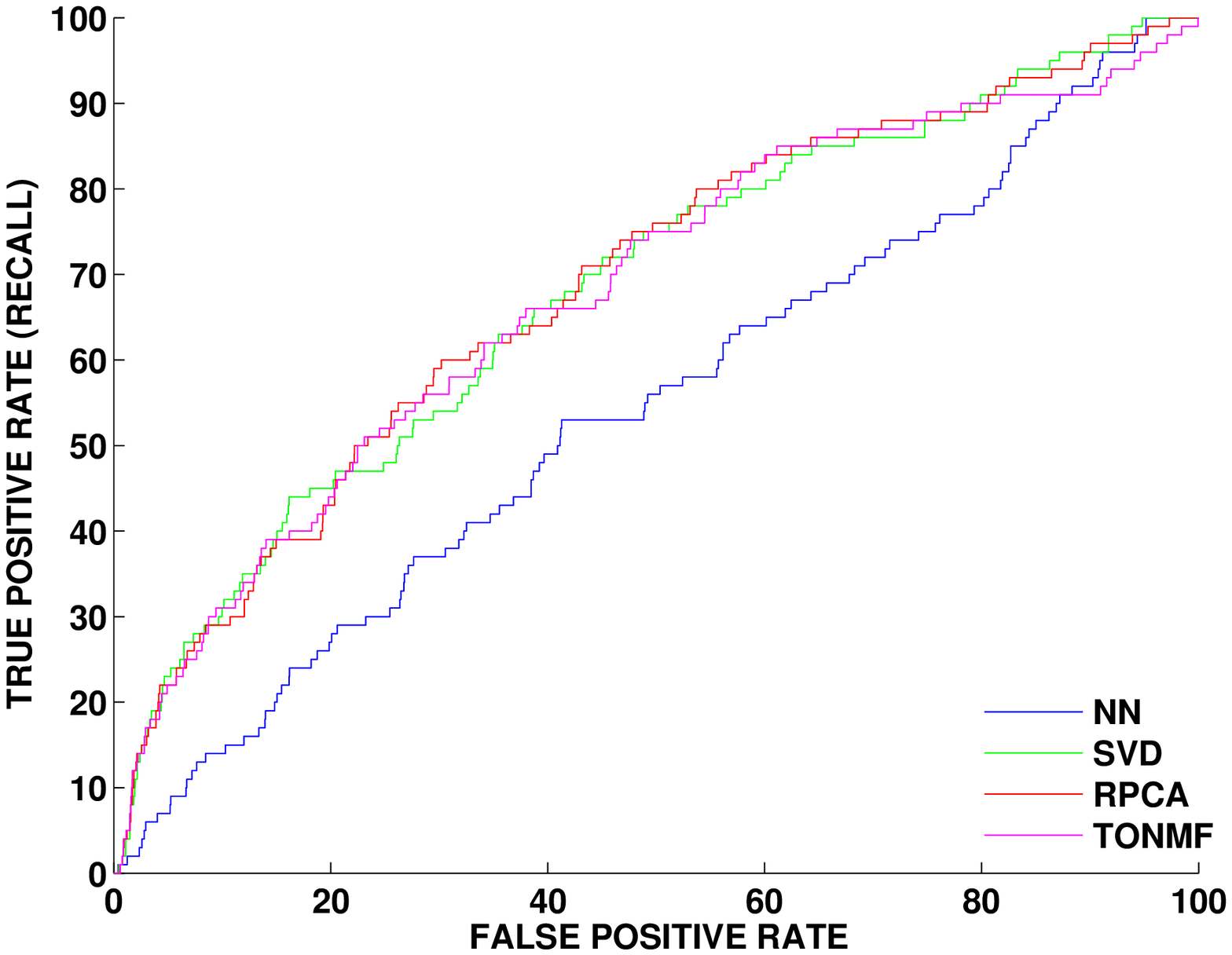}
  \caption{RCV20}
  \label{fig:rocrcv}
 \end{minipage}%
 \begin{minipage}{0.5\linewidth}
  \centering
  \caption*{Parameter Sensitivity}
  \includegraphics[width=2.4in,height=1.8in]{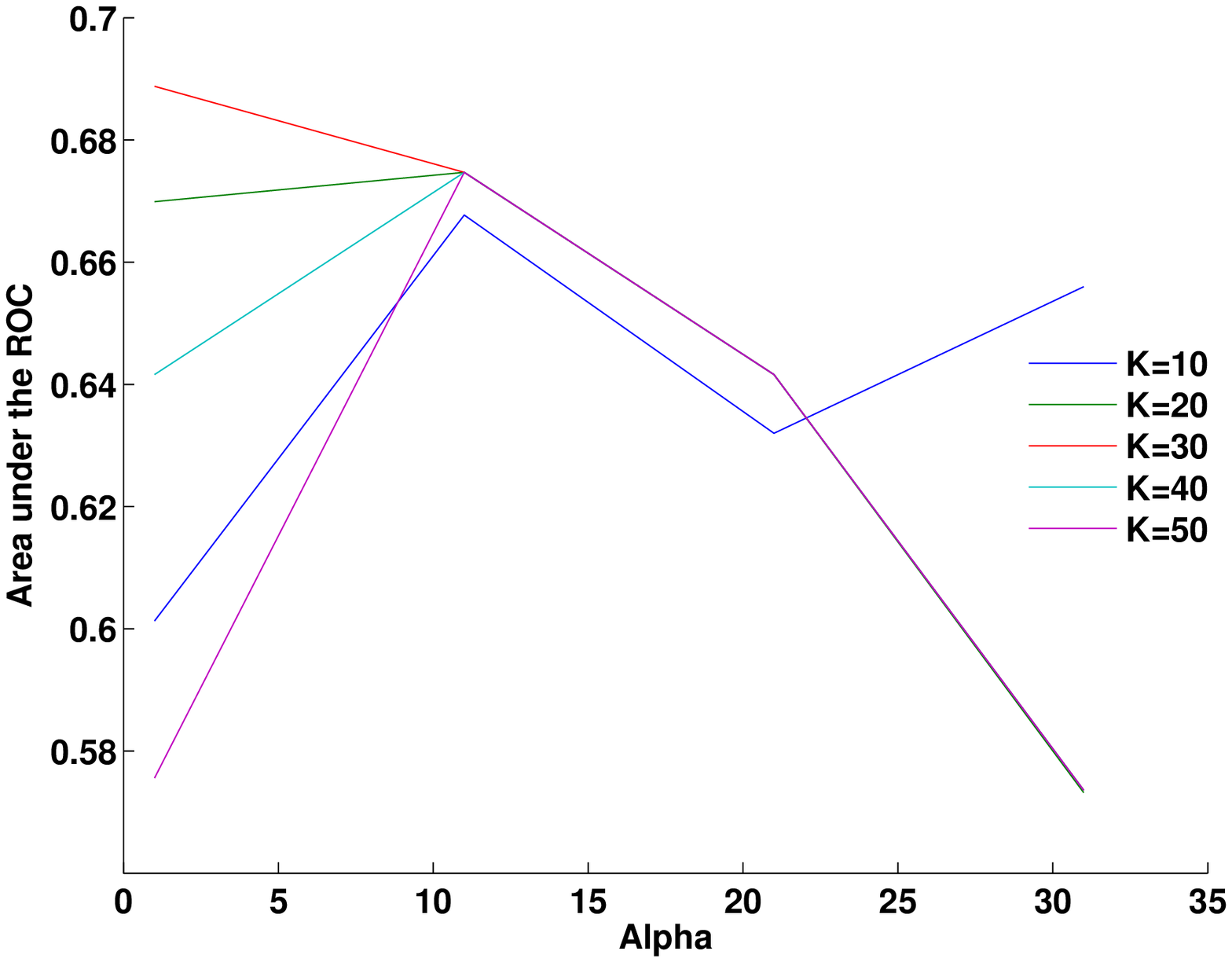}
  \caption{RCV20}
  \label{fig:alphakrcv}
 \end{minipage}
  \begin{minipage}{0.5\linewidth}
  \centering
  \caption*{ROC}
  \includegraphics[width=2.4in,height=1.8in]{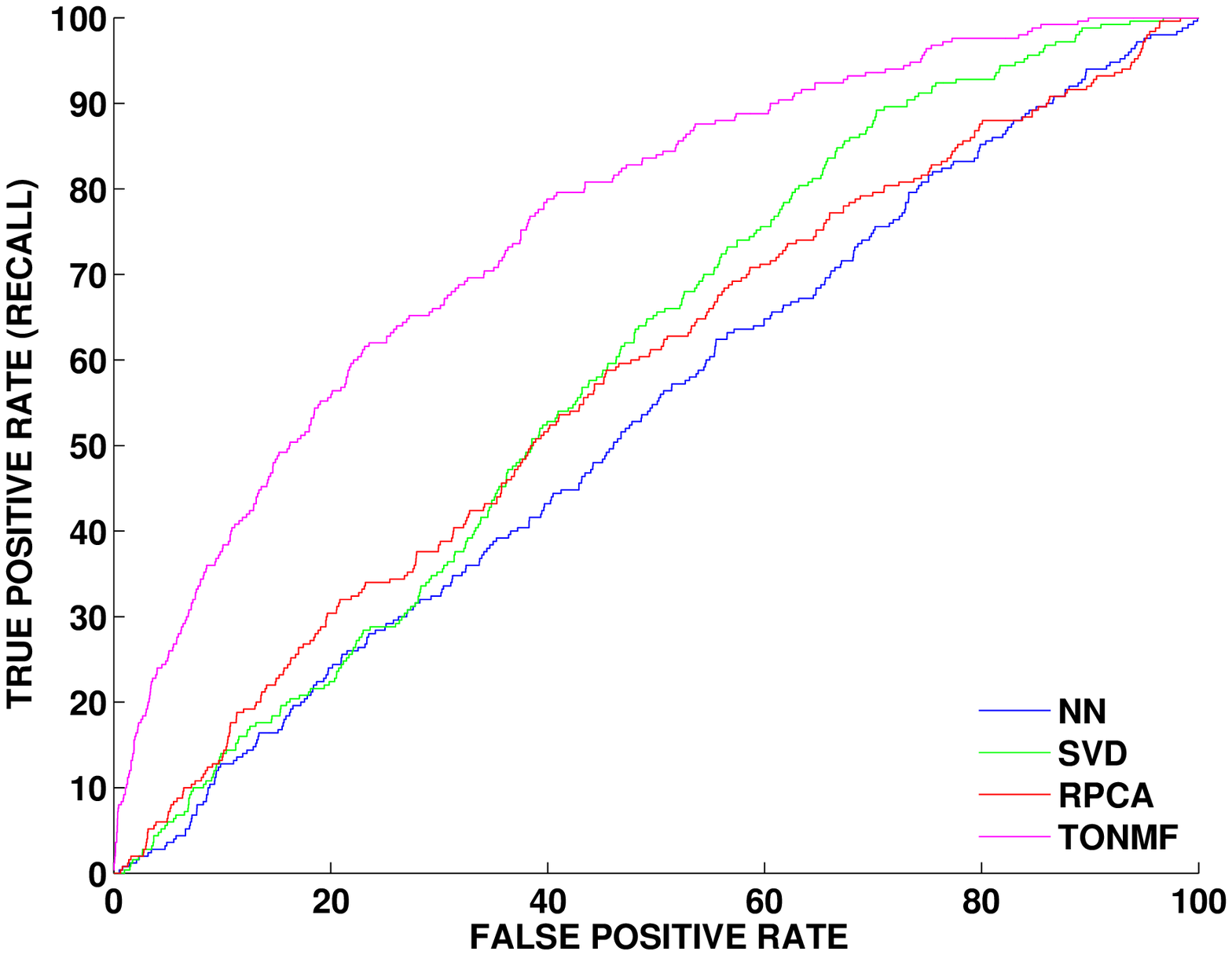}
  \caption{Market Basket}
  \label{fig:rocmb}
 \end{minipage}%
 \begin{minipage}{0.5\linewidth}
  \centering
  \caption*{Parameter Sensitivity}
  \includegraphics[width=2.4in,height=1.8in]{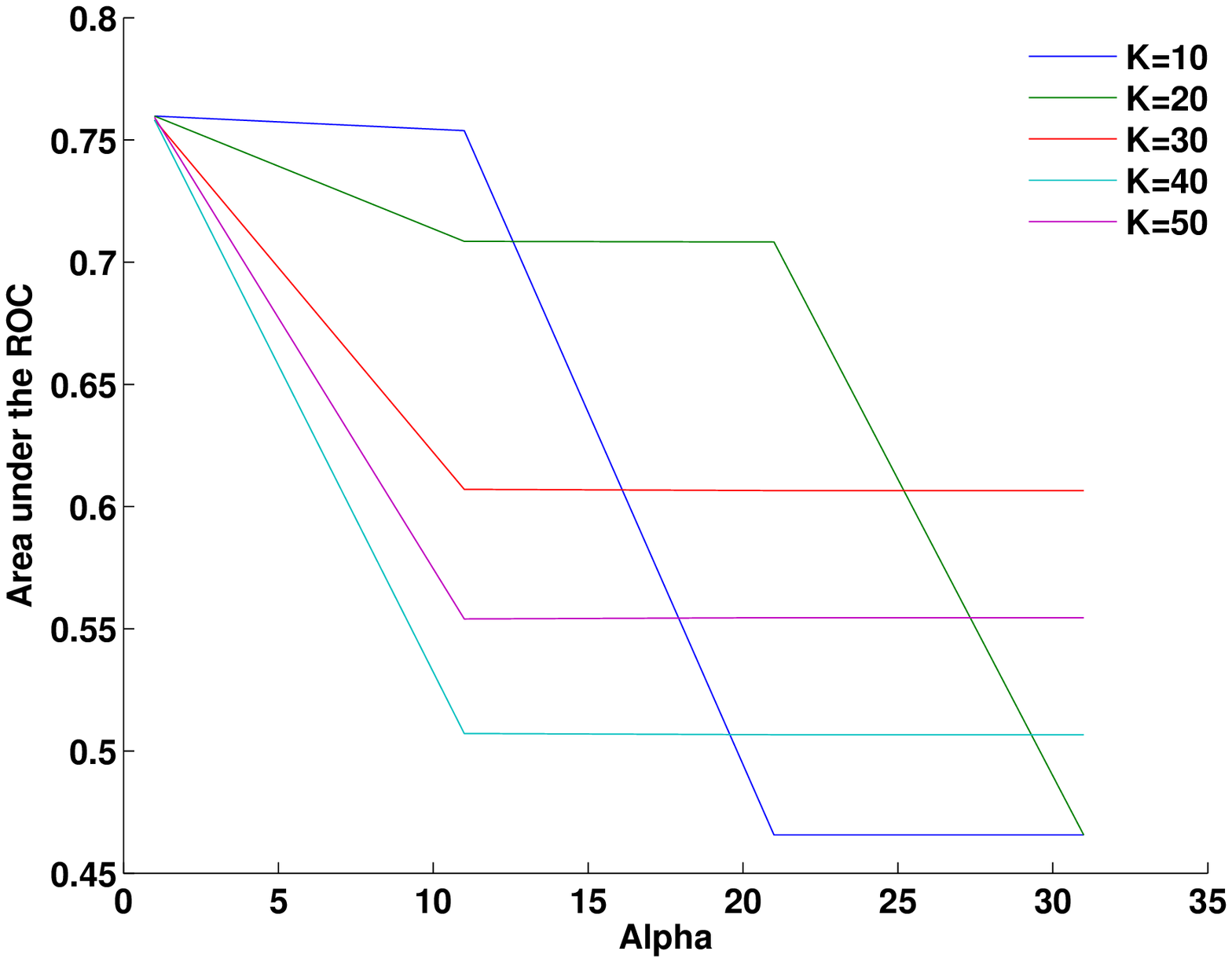}
  \caption{Market Basket}
  \label{fig:alphakmb}
 \end{minipage}
  \begin{minipage}{0.5\linewidth}
  \centering
  \caption*{ROC}
  \includegraphics[width=2.4in,height=1.8in]{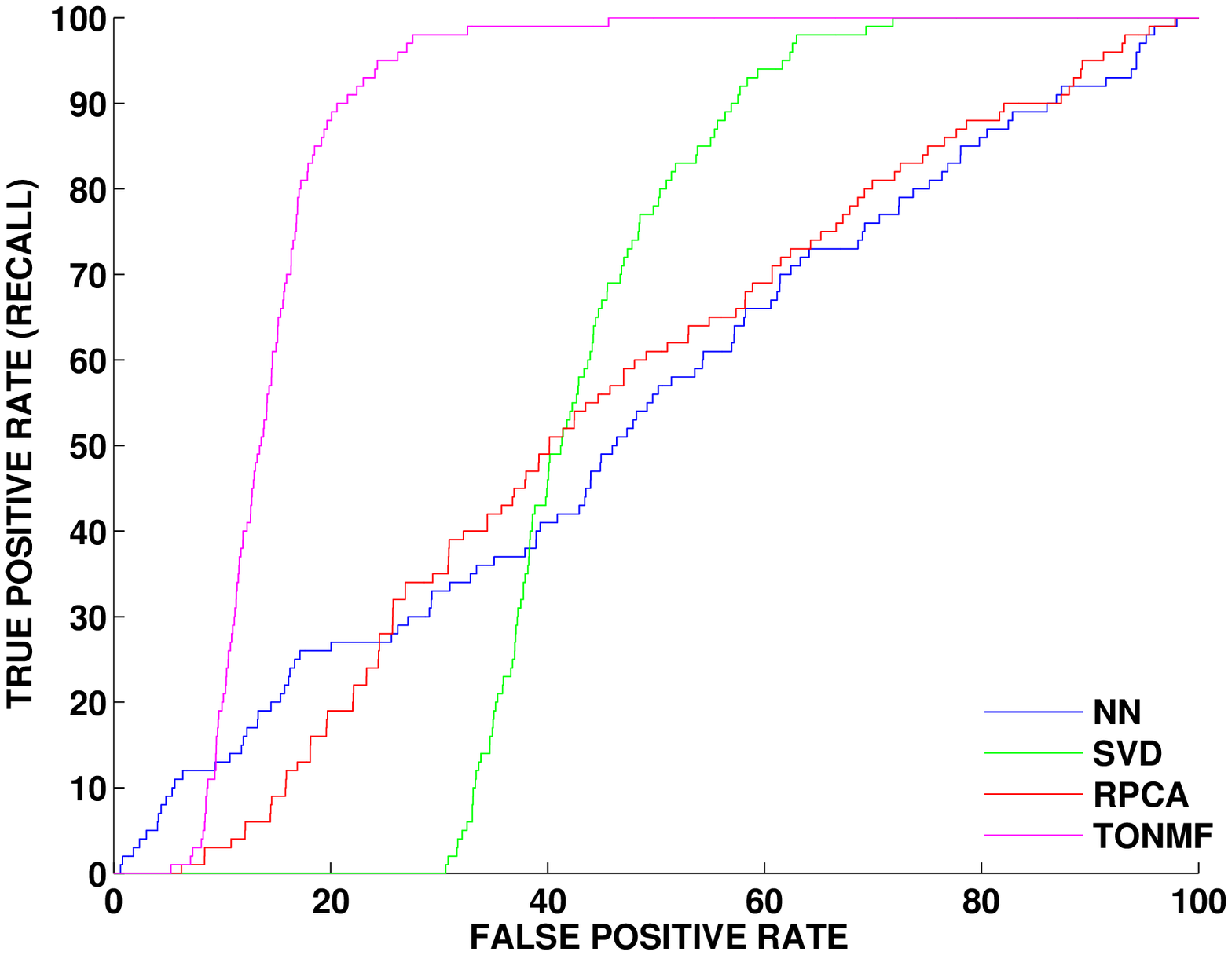}
  \caption{Wiki People}
  \label{fig:rocwiki}
 \end{minipage}%
 \begin{minipage}{0.5\linewidth}
  \centering
  \caption*{Parameter Sensitivity}
  \includegraphics[width=2.4in,height=1.8in]{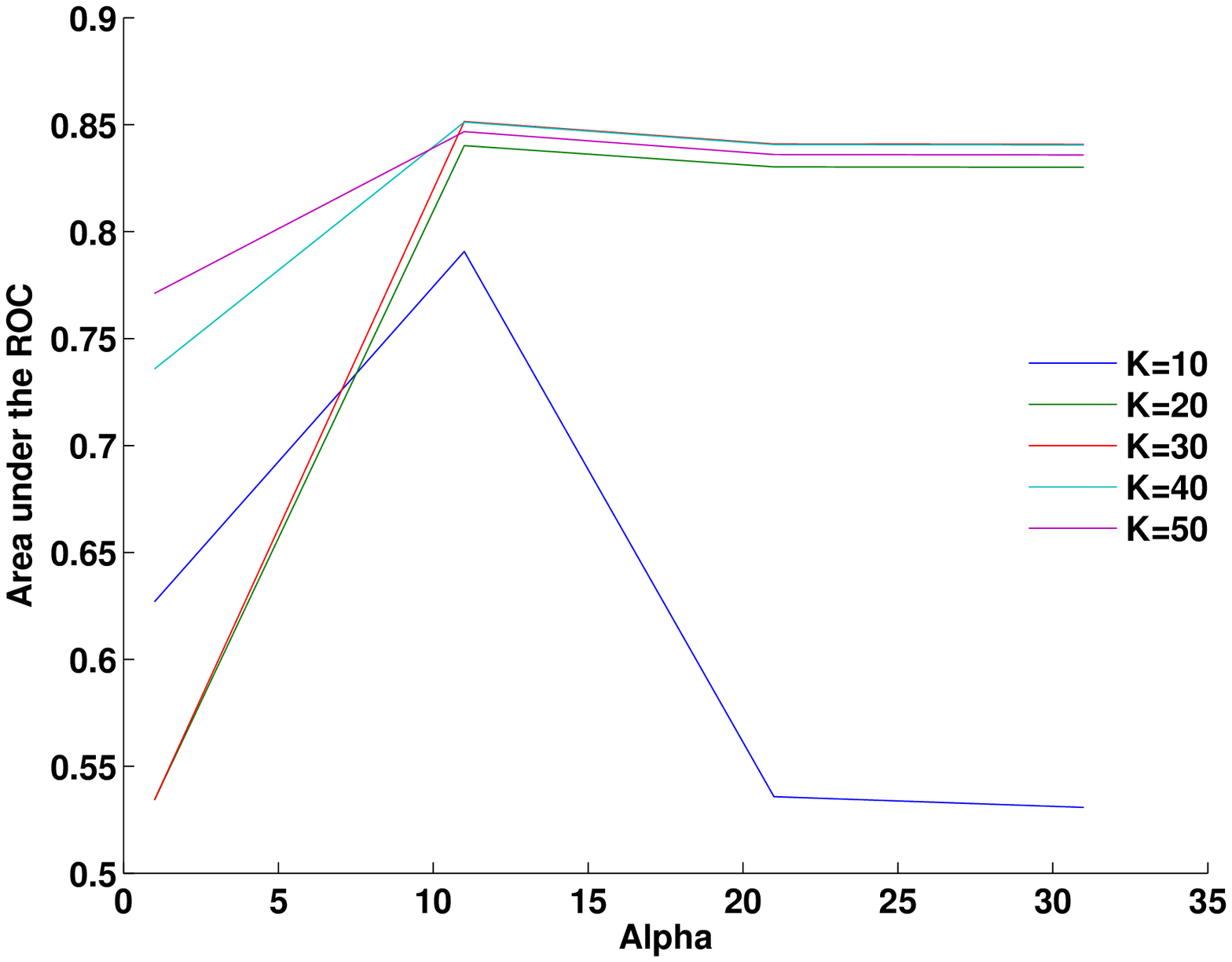}
  \caption{Wiki People}
  \label{fig:alphakwiki}
 \end{minipage}
\end{figure*}

We first present the ROC curves for the different data sets. The ROC
curve for the {\em Reuters} dataset is illustrated in Figure
\ref{fig:rocreuters}. In this case, our algorithm shows a drastic
improvement over both the baseline algorithms.  This is evident from
the rather large lift in the chart.   Our algorithm
\algo\, had an area of 0.9340 under ROC. The $k$-NN approach
performed quite poorly, and had an area under the  ROC curve of
0.5370. This is slightly better than random performance. The area
under ROC for the  $SVD$ method  was  0.5816 and \ramki{$RPCA$ was 0.6120}, which is better than
the $k$-NN method, but still significantly less than the proposed
algorithm.

 The comparison of our algorithm with baselines  for the
{\em RCV20} data set is shown in Figure \ref{fig:rocrcv}. As
discussed in the data generation section, this is a particularly
challenging data set, because of the similarity in the vocabulary
distribution between the rare class, and the regular class. It is
evident that our algorithm \algo\  performed better than the
$SVD$, $RPCA$  and the $k$-NN method. However, the lift in  the ROC curve for
all the methods is not particularly significant, because of the
inherently challenging nature of the data set. The $k$-NN
method performed particularly poorly in this case. In a later
section, we will provide some insights about the fact that some of
this ``poor'' performance is because of the noise in the  data set
itself, where some of the  points in the regular class should really
be considered outliers. We generated a datasets in RCV20 where we just changed the outlier class to {\em christian religion}. We received
a best ROC of 0.9732 and it is not shown in Figure \ref{fig:rocrcv}.

\ramki{ Figure \ref{fig:rocwiki} shows the comparison of our
algorithm $TONMF$ against the baselines for the {\em Wiki People}
data set.  The area under the ROC for $k$-NN was 0.5395, which is
rather poor. All the other methods performed better than $k$-NN with
area under the ROC for $SVD$ being 0.5670 and $RPCA$ being 0.5471.
Our algorithm $TONMF$ performed significantly better than all the
methods with an AUC of 0.8552. Clearly, this is a significant
qualitative difference between the methods. The above three were
experiments on real life dataset and we chose market basket for
synthetic dataset.}

 The ROC comparison for the synthetic  market basket data is illustrated
in Figure \ref{fig:rocmb}. In this case, the improvement of the
algorithm \algo\  over the baseline methods was quite  significant.
Specifically, the algorithm \algo\ had an area under the ROC curve
of 0.7598, which is a significant  lift. This significantly
outperformed the $SVD$ and \ramki{$RPCA$ method, which had an area under the ROC curve
of 0.5731 and 0.5758 respectively}. As in the case of the other data sets, the $k$-NN
algorithm performed very poorly with an area under the ROC curve of
0.5431. The consistently  poor performance of the $k$-NN approach
over all algorithms is quite striking, and suggests that
straightforward generalizations of  outlier analysis techniques from
other data domains are often not well suited to the text domain.

\ramki{Based on our conducted experiments on real world and
synthetic datasets, we observed that $TONMF$ outperformed every
other baseline.  Furthermore, the rank of the methods from  best  to
worst is $TONMF, RPCA, SVD$ and $NN$.} Clearly, conventional
distance-based  methods do not seem to work very well for text data.

\subsection{Parameter Sensitivity}
From (\ref{outlier}) in Section \ref{sec:model}, we can
see that the parameters for our algorithm are $\alpha, \beta$ and
the low rank $r$. We tested the algorithm for different variations
in the parameters,  and found that our algorithm was insensitive to
changes in $\beta$. In other words,  for a given low rank $r$ and
$\alpha$, the changes in the value of  $\beta$ did not result in
significant change in the area under ROC. Hence, in this paper, we
provide the charts of the ROC area  variation with the parameters
$\alpha$ and $r$ on the data sets.

The  sensitivity results for the {\em Reuters} data set are
illustrated in Figure \ref{fig:alphakreuters}. The value of $\alpha$
is illustrated on the $X$-axis, and different values of the low rank
$r$ are graphed by different curves in the plot. It is evident in
this case, that the
 area under the  ROC increased with
increase in low rank $r$ and $\alpha$. However the improvement
started diminishing and changed very marginally at higher ranks $r$.

The results for the {\em RCV20} and \ramki{{\em Wiki People}} datasets are illustrated in Figure \ref{fig:alphakrcv} \ramki{and Figure \ref{fig:alphakwiki} respectively}.
 As in the previous case, the value of $\alpha$ is illustrated on
 the $X$-axis, and  different values of the low rank $r$ are
 represented by different curves. In this case, the area under the
 ROC curve was relatively insensitive  to the parameters. This
 implies that the algorithm can be  used over a wide range of
 parameters, without affecting the performance too much.
Finally, the results for the market basket data set are illustrated
in  Figure \ref{fig:alphakmb}.  In this case, the area under the ROC
curve decreases with increase in low rank $r$ and $\alpha$. This is
because the market-basket data  has inherently very low (implicit)
dimensionality, and  therefore, it is best to use  a relatively low
rank in order to mine the outliers.

\ramki{From the parameter sensitivity graphs for real world
datasets, we observe that for a given $\alpha$, the approach is
relatively insensitive to the rank of the approximation. It needs to
be kept in mind that it is generally faster to determine approximations
with lower rank.  This implies that, for very large matrices, the
algorithm can be made computationally faster by choosing approximations 
with lower rank without compromising on the performance. According to
the model explained in equation \eqref{outlier}, the parameters
$\alpha$ and $\beta$ balance the importance given to outliers
against the  matrix sparsity criterion during regularization. By
picking $\alpha
>> \beta$, the importance of the outlier portion of the regularization
increases. From the parameter sensitivity graph,  it is evident that
for most low ranks $K$, the increase in the value of $\alpha$ does
not improve the performance of the outlier detection. This is
because, beyond a particular limit, the weights given to  the
outlier criterion  do not supersede the optimization problem's main
objective of extracting the low-rank patterns from the underlying
data. }

\subsection{Further Insights} \label{sec:insights}
In order to illustrate the inner workings of  the matrix
factorization approach, we provide some further insights about the
statistics buried deep in the algorithm. We also present some interesting
observations when outliers share the same vocabulary distribution as
regular data points, as is the case for the {\em RCV20} data set.
One observation is that the method of data generation implicitly
assumes that all the documents within a ``regular'' class in a real
data set are not outliers. This is of course not true in practice,
since some of the documents within these classes will also be
outliers, for reasons other than topical affinity.  Our algorithm
\algo\ was also able to  detect such distinct documents, much better
than the other baseline algorithms. We isolated those false
positives of our algorithm \algo\ that were not detected in the
baselines in the case of the {\em RCV20} data set. It was observed
that while these outliers officially belonged to one of the regular
classes, they did show different {\em kinds} of distinctive
characteristics. For example, while the average number of words in
regular documents was 195, the ``false positive'' outliers chosen by
our algorithm were typically either very lengthy with  over 400
words, or were unusually short will less than 150 words.  This
behaviour was also generally reflected in the number of distinct
words per document. Another observation is that these outlier
documents typically had a significant vocabulary repetition over a
small number of distinct words. Thus,  the algorithm was also able
to identify those natural outliers, which {\em ought to} have been
considered outliers for reasons of statistical word distribution, as
opposed to their topical behaviour.

% \begin{figure}[ht]
% \centering
% \subfigure[Neutral Smiley]{%
 % \includegraphics[width=3.2in,height=2in]{rocmbsmallcomp.eps}.
% \label{fig:subfigure1}}
% \quad
% \subfigure[Blush Smiley]{%
% \includegraphics[width=3.2in,height=2in]{alphakmbsmall.eps}
% \label{fig:subfigure2}}
% \subfigure[Sleepy Smiley]{%
% \includegraphics[width=3.2in,height=2in]{rocrcv.eps} .
% \label{fig:subfigure3}}
% \quad
% \subfigure[Angry Smiley]{%
% \includegraphics[width=3.2in,height=2in]{alphakrcv.eps}
% \label{fig:subfigure4}}
% \subfigure[c]{
% \includegraphics[width=3.2in,height=2in]{rocreuters.eps}
% \label{c}}
% \quad
% \subfigure[d]{
% \includegraphics[width=3.2in,height=2in]{alphakreuters.eps}
% \label{d}}
% \caption{Main figure caption}
% \label{fig:figure}
% \end{figure}

\section{Conclusion}
\label{sec:conclusion}

This paper presents a matrix factorization based approach to text
outlier analysis. The approach is designed to adjust well to the
widely varying structures in different localities of the data, and
therefore provides more robust methods than competing models. The
approach has the potential to be applied to other domains with
similar structure, and as a specific example, we provide experiments
on market  basket data. We also presented extensive experimental
results, which illustrate the superiority of the approach.  
Our code can be downloaded from 
\url{https://github.com/ramkikannan/outliernmf} and 
tried with any text dataset. 

In this paper, we had a parallel implementation using the
Matlab's parallel computing toolbox to run in multicore environments.
In the future, we would like to explore a scalable implementation
of our algorithm. The solution is embarrassingly parallelizable,
and would like to experiment in web scale data. One of the potential
extension is incorporating temporal and spatial aspects into the model.
Such an extension, make the solution applicable to emerging 
applications such as topic detection and streaming data. 
%In the recent times,
%approximate matrix factorization techniques are explored by
%randomly sampling the input matrix. We can reduce the computation
%time for very large matrices using such sampling techniques. We would like
%to explore a sampling based solution for our model. 
We experimented
the solution primarily on text data and market basket data. In future
work, we will extend this broader approach to other domains such as
video data.

\section{Acknowledgements}

This manuscript has been co-authored by UT-Battelle, LLC under Contract No. DE-AC05-00OR22725 with the U.S. Department of Energy.  This project was partially funded by the Laboratory Director's Research and Development fund and also sponsored by the Army Research Laboratory (ARL) and was accomplished under Cooperative Agreement Number W911NF-09-2-0053. Also, H. Woo is supported by NRF-2015R101A1A01061261. 

%This research used resources of the Oak Ridge Leadership Computing Facility at the Oak Ridge National Laboratory, which is supported by the Office of Science of the U.S. Department of Energy.

The United States Government retains and the publisher, by accepting the article for publication, acknowledges that the United States Government retains a non-exclusive, paid-up, irrevocable, world-wide license to publish or reproduce the published form of this manuscript, or allow others to do so, for United States Government purposes. The Department of Energy will provide public access to these results of federally sponsored research in accordance with the DOE Public Access Plan (\url{http://energy.gov/downloads/doepublic-access-plan}).

Any opinions, findings and conclusions or recommendations expressed in this material are those of the authors and do not necessarily reflect the views of the USDOE, NSF or ARL.

\bibliographystyle{abbrv}
\bibliography{references}

\end{document}